\documentclass[final,noappendix, runningheads]{llncs}
\usepackage{wrapfig}
\usepackage{listings}
\usepackage{amssymb}
\usepackage{amsmath}
\usepackage{amsfonts}
\usepackage{mathpartir}
\usepackage[inference]{semantic}
\usepackage{stackengine}
\usepackage{xcolor}
\stackMath
\usepackage{ifthen}
\usepackage{paralist}
\usepackage{hyperref}
\usepackage{xspace}
\usepackage{multirow}
\usepackage{tikz}
\usetikzlibrary{automata, positioning, arrows.meta, arrows, decorations.pathmorphing,shapes,tikzmark}

\pgfarrowsdeclare{biggertip}{biggertip}{  
  \setlength{\arrowsize}{0.4pt}  
  \addtolength{\arrowsize}{.5\pgflinewidth}  
  \pgfarrowsrightextend{0}  
  \pgfarrowsleftextend{-5\arrowsize}  
}{  
  \setlength{\arrowsize}{0.4pt}  
  \addtolength{\arrowsize}{.5\pgflinewidth}  
  \pgfpathmoveto{\pgfpoint{-5\arrowsize}{4\arrowsize}}  
  \pgfpathlineto{\pgfpointorigin}  
  \pgfpathlineto{\pgfpoint{-5\arrowsize}{-4\arrowsize}}  
  \pgfusepathqstroke  
}  
\usepackage[normalem]{ulem}

\usepackage{alltt}

\usepackage{pgfplots}
\pgfplotsset{compat=1.16}
\usepackage{varwidth}
\usepackage{relsize}
\tikzset{
	->, %
	>=Stealth, %
	node distance=3cm, %
	every state/.style={thick, fill=gray!10}, %
	initial text=$ $, %
}

\usepackage[textsize=footnotesize,obeyFinal]{todonotes} %

\usepackage{centernot}
\usepackage{upgreek}
\usepackage{rotating}
\usepackage[linesnumbered,ruled,vlined]{algorithm2e}
\SetKwComment{Comment}{/* }{ */}

\SetCommentSty{mycommfont}\usepackage{mathtools}

\usepackage[appendix=append,bibliography=common]{apxproof}

\newtheoremrep{apxtheorem}{Theorem}
\newtheoremrep{apxlemma}{Lemma}
\newtheoremrep{apxproposition}{Proposition}
\newtheoremrep{apxcorollary}{Corollary}

\newcommand\pfun{\mathrel{\ooalign{\hfil$\mapstochar\mkern5mu$\hfil\cr$\to$\cr}}}

\usepackage{graphicx}

\newcounter{sqindex}

\newcommand{\np}[1]{\todo[color=cyan]{NP: {#1}}}

\newcommand{\lds}[1]{\todo[color=orange]{LDS: {#1}}}

\newtheoremrep{theorem}{Theorem}
\newtheoremrep{lemma}{Lemma}
\newtheoremrep{proposition}{Proposition}

\makeatletter %
\def\arcr{\@arraycr}
\makeatother

\makeatletter
\newcommand{\xdashrightarrow}[2][]{\ext@arrow 0359\rightarrowfill@@{#1}{#2}}
\newcommand{\xdashleftarrow}[2][]{\ext@arrow 3095\leftarrowfill@@{#1}{#2}}
\newcommand{\xdashleftrightarrow}[2][]{\ext@arrow 3359\leftrightarrowfill@@{#1}{#2}}
\def\rightarrowfill@@{\arrowfill@@\relax\relbar\rightarrow}
\def\leftarrowfill@@{\arrowfill@@\leftarrow\relbar\relax}
\def\leftrightarrowfill@@{\arrowfill@@\leftarrow\relbar\rightarrow}
\def\arrowfill@@#1#2#3#4{%
	$\m@th\thickmuskip0mu\medmuskip\thickmuskip\thinmuskip\thickmuskip
	\relax#4#1
	\xleaders\hbox{$#4#2$}\hfill
	#3$%
}
\makeatother

\newcommand\mydef{\mathrel{\overset{\makebox[0pt]{\mbox{\normalfont\tiny\sffamily def}}}{=}}}
\newcommand\ttt{\textbf{tt}}
\newcommand\fff{\textbf{ff}}

\def\inc{{\hspace{-.04em}\raisebox{.25ex}{\tiny\bf ++}}}
\def\dec{{\hspace{-.01em}\raisebox{.25ex}{\tiny\bf \textminus\textminus}}}

\def\cbigwedge{\bigwedge\mspace{-15mu}\bigwedge}

\def\Cs{\textit{Cs}}

\newcommand{\prop}{\Sigma}
\newcommand{\arena}{A\xspace}
\newcommand{\ltlprop}{\mathbb{AP}\xspace}
\renewcommand{\date}{{DATE}\xspace}

\newcommand\env{\mathbb{E}}
\newcommand\con{\mathbb{C}}

\newcommand\preds{\mathcal{P}r}

\newcommand\counterstrategy{{counterstrategy}\xspace}
\newcommand\Counterstrategy{{Counterstrategy}\xspace}
\newcommand\counterstrategies{{counterstrategies}\xspace}

\newcommand\mealy{\ensuremath{\mathit{MM}}\xspace}
\newcommand\moore{\ensuremath{\mathit{Cs}}\xspace}

\def\inc{{\hspace{-.04em}\raisebox{.25ex}{\tiny\bf ++}}}

\newcommand\ltl[1][]{%
\ifthenelse{\equal{#1}{}}
  {{\textrm{LTL}}}
  {{\textrm{LTL}$(#1)$}}%
\xspace
}

\newcommand\abs[2][]{%
\ifthenelse{\equal{#1}{}}
  {{\alpha({#2})}}
  {{\alpha_{_{#1}}({#2})}}
}

\newcommand{\pair}[1]{{\langle {#1} \rangle}}

\newcounter{sarrow}

\newcommand\false{\textit{false}\xspace}
\newcommand\true{\textit{true}\xspace}

\definecolor{mGreen}{rgb}{0,0.6,0}
\definecolor{mGray}{rgb}{0.5,0.5,0.5}
\definecolor{mPurple}{rgb}{0.58,0,0.82}
\definecolor{backgroundColour}{rgb}{0.95,0.95,0.92}

\lstdefinestyle{CStyle}{
    backgroundcolor=\color{backgroundColour},   
    commentstyle=\color{mGreen},
    keywordstyle=\color{magenta},
    numberstyle=\tiny\color{mGray},
    stringstyle=\color{mPurple},
    basicstyle=\footnotesize,
    breakatwhitespace=false,         
    breaklines=true,                 
    captionpos=b,                    
    keepspaces=true,                 
    numbers=left,                    
    numbersep=5pt,                  
    showspaces=false,                
    showstringspaces=false,
    showtabs=false,                  
    tabsize=2,
    language=C
}

\newcommand{\citet}[1]{\cite{#1}}
\newcommand{\val}{\textit{val}}
\newcommand{\vals}{\textit{vals}}
\newcommand{\Val}{\textit{Val}}

\newcommand{\sweap}{S\xspace}

\usepackage{graphicx}
\usepackage{paralist}

\usepackage{pifont} %
\usepackage[caption=false]{subfig}
\usepackage[firstpage,stamp]{draftwatermark} %
\usepackage[misc]{ifsym} %

\SetWatermarkAngle{0}
\SetWatermarkText{\raisebox{12.5cm}{
\href{https://doi.org/10.5281/zenodo.15129663}{\includegraphics{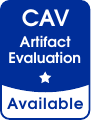}}
\hspace{9cm}
\includegraphics{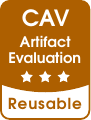}
}}

\begin{document}
\title{Full LTL Synthesis over Infinite-state Arenas\thanks{This work is funded by
		the ERC consolidator grant
		D-SynMA (No. 772459) and the Swedish research council
		project (No. 2020-04963).}}

\author{
Shaun Azzopardi\inst{3}\textsuperscript{(\Letter)}\orcidID{0000-0002-2165-3698} \and
Luca Di Stefano\inst{1,2}\orcidID{0000-0003-1922-3151} \and
Nir Piterman\inst{1}\orcidID{0000-0002-8242-5357} \and
Gerardo Schneider\inst{1}\orcidID{0000-0003-0629-6853}
}
\authorrunning{Azzopardi et al.}

\institute{
University of Gothenburg and Chalmers University of\\ Technology, Gothenburg, Sweden \and
TU Wien, Institute of Computer Engineering,\\Treitlstraße 3,1040 Vienna, Austria\\\and
Dedaub, San Gwann, Malta
\email{shaun.azzopardi@gmail.com}
}
\maketitle              %

\begin{abstract}
Recently, interest has increased in applying reactive synthesis to richer-than-Boolean domains.
A major (undecidable) challenge in this area is to establish when certain repeating behaviour terminates in a desired state when the number of steps is unbounded. 
Existing approaches struggle with this problem, or can handle at most deterministic games with B{\"u}chi goals.
This work goes beyond by contributing the first effectual approach to synthesis with full LTL objectives, based on Boolean abstractions that encode both safety and liveness properties of the underlying infinite arena.
We take a CEGAR approach: attempting synthesis on the Boolean abstraction, checking spuriousness of abstract \counterstrategies through invariant checking, and refining the abstraction based on counterexamples.
We reduce the complexity, when restricted to predicates, of abstracting and synthesising by an exponential through an efficient binary encoding. This also allows us to eagerly identify useful fairness properties.
Our discrete synthesis tool outperforms the state-of-the-art on linear integer arithmetic (LIA) benchmarks from literature,
solving almost double as many syntesis problems as the current state-of-the-art. It also solves slightly more problems than the second-best realisability checker, in one-third of the time.
We also introduce benchmarks with richer objectives that other approaches cannot handle, and evaluate our tool on them.

\end{abstract}
\section{Introduction}
\label{sec:intro}

Reactive synthesis provides a way to synthesise controllers that ensure satisfaction of high-level \emph{Linear Temporal Logic} (LTL) specifications, against uncontrolled environment behaviour. 
Classically, synthesis was suggested and applied in the Boolean (or finite-range) variable setting~\cite{DBLP:conf/popl/PnueliR89}.
Interest in the infinite-range variable setting was soon to follow.
Some of the milestones include the adaptation of the theory of CEGAR to infinite-state games~\cite{DBLP:conf/icalp/HenzingerJM03} and the early adoption of SMT for symbolic representation of infinite-sized sets of game configurations~\cite{DBLP:conf/popl/BeyeneCPR14}.
However, in recent years, success of synthesis in the finite domain as well as maturity of SMT solvers has led to sharply growing interest in synthesis in the context of infinite-range variables, with several tools becoming available that tackle this problem. 
We highlight the two different (but related) approaches taken by the community:
(a)
application of infinite-state reactive synthesis from extensions of LTL where atoms include quantifier-free first-order formulas over infinite-range variables~\cite{MaderbacherBloem22,10.1145/3519939.3523429,10.1007/978-3-030-25540-4_35,DBLP:conf/isola/MaderbacherWB24} and 
(b) direct applications to the solution of games with an infinite number of configurations~\cite{10.1007/978-3-030-81685-8_42,10.1145/3632899,DBLP:conf/cav/SchmuckHDN24,DBLP:journals/pacmpl/HeimD25}.
Two notable examples of the two approaches from the last two years include:
(a) the identification of a fragment of LTL with first-order atoms that allows
 for a decidable synthesis framework~\cite{DBLP:conf/cav/RodriguezS23,DBLP:journals/jlap/RodriguezS24,DBLP:conf/aaai/Rodriguez024} and (b) the introduction of so-called \emph{acceleration lemmas}~\cite{10.1145/3632899,DBLP:conf/cav/SchmuckHDN24,DBLP:journals/pacmpl/HeimD25} targeting the general undecidable infinite-state synthesis problem.
The latter directly attacks a core issue of the problem's undecidability: identify whether certain repeated behaviour can eventually force the interaction to a certain state.
Thus, solving the (alternating) termination problem.

Infinite-state reactive synthesis aims at producing a system that manipulates variables with infinite domains and reacts to input variables controlled by an adversarial environment. 
Given an LTL objective, the \emph{realisability problem} is to determine whether a system may exist that enforces the objective.
Then, the \emph{synthesis problem} is to construct such a system, or a \emph{counterstrategy} by which the environment may enforce the negation of the objective.
While in the finite-state domain realisability and synthesis are tightly connected, this is not the case in the infinite-state domain and many approaches struggle to (practically) scale from realisability to synthesis.
In this paper we focus on the more challenging synthesis problem, rather than mere realisability, to be able to construct implementations. \np{Can we mention some connection bewteen practical applicability and our tool?}
Furthermore, our approach is tailored for the general~--~undecidable~--~case.

As mentioned, a major challenge is the identification of repeated behaviour that forces reaching a given state.
Most approaches rely on one of two basic techniques: 
either refine an abstraction based on a mismatch in the application of a transition between concrete and abstract
representations, or compute a representation of the set of immediate successors/predecessors of a given set of states.
Both have limited effectiveness due to the termination challenge.
Indeed, in many interesting cases, such approaches attempt at enumerating paths of unbounded length.
For example, this is what happens to approaches relying on refinement~\cite{MaderbacherBloem22,10.1007/978-3-030-25540-4_35}, which is sound but often cannot terminate. 
It follows that reasoning about the effect of repeated behaviour is crucial.

We know of two attempts at such reasoning.
\texttt{temos}~\cite{10.1145/3519939.3523429} identifies single-action loops that terminate in a desired state,
but cannot generalise to more challenging cases, e.g., where the environment may momentarily interrupt the loop,
and moreover it cannot supply unrealisability verdicts.
By contrast,
\texttt{rpgsolve}~\cite{10.1145/3632899}
summarises terminating sub-games via acceleration lemmas to construct an argument for realisability, relying on quantifier elimination with uninterpreted functions. 
However, this approach is limited to at most deterministic B{\"u}chi objectives, and 
is practically more effective for realisability than for synthesis due to the challenges of quantifier elimination. 
Its extension \texttt{rpg-STeLA}~\cite{DBLP:conf/cav/SchmuckHDN24} attempts to identify  acceleration lemmas that apply to multiple regions and thus solves games compositionally, but only supports realisability. 

\begin{figure}[t]
    \centering
    \includegraphics[width=0.9\columnwidth]{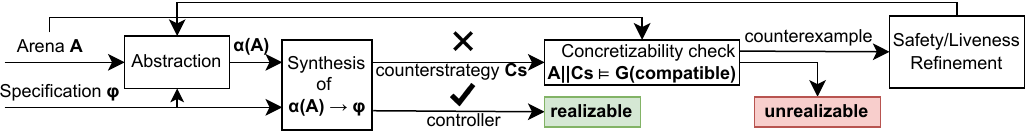}
    \caption{Workflow of our approach.}
    \label{fig:approach}
\end{figure}

In this paper we address the limitations described above, generalising infinite-state reactive synthesis to more expressive objectives.
In particular, we consider LTL objectives over infinite-state arenas, without imposing any limit on temporal nesting. 
Similar to others, our atoms may include quantifier-free first-order formulas.
However, we do not restrict the LTL formulas.
Furthermore, our approach does not distinguish between realisability and synthesis, and can synthesise both controllers and \counterstrategies. 
As shown in Fig.~\ref{fig:approach}, our approach is based on CEGAR~\cite{DBLP:conf/popl/HenzingerJMS02}, heavily adapted for synthesis.
Our main contributions are:
\begin{compactenum}
    \item An efficient binary encoding of predicates. This reduces complexity, in terms of predicates, of abstraction building/size from exponential to polynomial, and of finite synthesis over abstractions from doubly to singly exponential.
    \item A method to check \counterstrategy concretisability through invariant checking, that finds minimal counterexamples to concretisability.
    \item Two new kinds of liveness refinements:
    \emph{Structural refinement}, which monitors for terminating concrete loops in the abstract system, and enforces eventual exit; and \emph{Ranking refinement} that relies on the binary encoding, which ensures the well-foundedness of terms relevant to the game in the abstraction. 
    \item An implementation of the above contributions for LIA problems. 
    \item The most extensive experimental comparison of infinite-state LIA realisability and synthesis tools in literature. This shows our tool substantially outperforming all others, making it the new state-of-the-art. 
    \item Separately, we enrich the dataset of existing benchmarks, which currently include at most weak fairness requirements, with a selection of problems incorporating strong fairness.
\end{compactenum}

For the reader's convenience we present the approach informally in Section~\ref{sec:infovr}, before formalising it in detail (Sections~\ref{sec:setting},~\ref{sec:abstoconc},~\ref{sec:ref}).
Then we describe our techniques to improve its efficiency (Section~\ref{sec:efficient}),
present and evaluate our tool (Section~\ref{sec:eval}),
and conclude while also discussing related and future work (Sections~\ref{sec:disc}--\ref{sec:conc}).

\section{Background}
\label{sec:background ltl}

We use the following notation throughout: for sets $S$ and $T$ such that $S\subseteq T$, we write $\cbigwedge_{T} S$ for $\bigwedge S \wedge \bigwedge_{s \in T \setminus S} \neg s$. We omit set $T$ when clear from the context. %

\smallskip\noindent
$\mathbb{B}(S)$ is the set of Boolean combinations of a set $S$ of Boolean variables.

\smallskip
\noindent
\emph{Linear Temporal Logic}, \ltl[\ltlprop], is the language over a set of propositions $\ltlprop$, defined as follows,\footnote{See \cite{DBLP:reference/mc/PitermanP18} for the standard semantics.} where $p \in \ltlprop$: $\phi \mydef \ttt \mid \fff \mid p \mid \neg \phi \mid \phi \wedge \phi \mid \phi \vee \phi \mid X\phi \mid \phi U \phi$.

\noindent 
For $w\in (2^{\ltlprop})^\omega$, we write $w \models \phi$ or $w \in L(\phi)$, when $w$ satisfies $\phi$.

\smallskip
\noindent
A \emph{Moore machine} is $C = \langle S, s_0, \prop_{in}, \prop_{out}, \rightarrow, out\rangle$, where $S$ is the set of states, $s_0$ the initial state, $\prop_{in}$ the set of input events, $\prop_{out}$ the set of output events, $\rightarrow :S \times 2^{\prop_{in}} \mapsto S$ the complete deterministic transition function, and $out: S \mapsto 2^{\prop_{out}}$ the labelling of each state with a set of output events. 
For $(s,I,s') \in \rightarrow$, where $out(s)=O$ we write $s \xrightarrow{I/O} s'$.

\smallskip
\noindent
A \emph{Mealy machine} is $C = \langle S, s_0, \prop_{in}, \prop_{out}, \rightarrow\rangle$, 
where $S$, $s_0$, $\prop_{in}$, and $\prop_{out}$ are as before and 
$\rightarrow : S \times 2^{\prop_{in}} \mapsto 2^{\prop_{out}} \times S$ the complete deterministic transition function. 
For $(s,I,O,s') \in \rightarrow$ we write $s \xrightarrow{I/O} s'$. 

Unless mentioned explicitly, both Mealy and Moore machines can have an infinite number of states.
A \emph{run} of a machine $C$ is $r=s_0,s_1,\ldots$ such that 
for every $i\geq 0$ we have $s_i \xrightarrow{I_i/O_i} s_{i+1}$ for some $I_i$ and $O_i$.
Run $r$ \emph{produces} the word $w=\sigma_0,\sigma_1,\ldots$, where $\sigma_i = I_i\cup O_i$.
A machine $C$ produces the word $w$ if there is a run $r$
producing $w$.
Let $L(C)$ denote the set of all words produced by $C$.

\smallskip
\noindent We cast our synthesis problem into the \emph{LTL reactive synthesis problem}, which calls for finding a Mealy machine that satisfies a given specification over input and output variables $\env$ and $\con$.
	
\begin{definition}[LTL Synthesis]
	A specification $\phi$ over $\env \cup \con$ is said to be \emph{realisable} if and only if there is a Mealy machine $C$, with input $2^\env$ and output $2^\con$, such that for every $w \in L(C)$ we have $w \models\phi$. We call $C$ a \emph{controller} for $\phi$.

    A specification $\phi$ is said to be \emph{unrealisable} if there is a Moore machine $\moore$, with input $2^\con$ and output $2^\env$, such that for every $w \in L(\moore)$ we have that $w \models \neg \phi$. We call $\moore$ a \emph{\counterstrategy} for $\phi$.

    The problem of synthesis is to construct $C$ or $\moore$, exactly one of which exists. 
\end{definition}

Note that the duality between the existence of a strategy and \counterstrategy follows from the determinacy of turn-based two-player $\omega$-regular games~\cite{10.2307/1971035}. 
We know that finite-state machines suffice for synthesis from LTL specifications~\cite{DBLP:conf/popl/PnueliR89}.

To be able to represent infinite synthesis problems succinctly we consider formulas in a theory.
A \emph{theory} consists of a set of terms and predicates over these. Atomic terms are constant values ($\mathcal{C}$) or variables. Terms can be constructed with operators over other terms, with a fixed interpretation. The set $\mathcal{T}(V)$ denotes the terms of the theory, with free variables in $V$. For $t \in \mathcal{T}(V)$, we write $t_{prev}$ for the term where variables $v$ appearing in $t$ are replaced by fresh variables $v_{prev}$. 

We use $\mathcal{ST}(V)$ to denote the set of \emph{state predicates}, i.e., predicates over $\mathcal{T}(V)$, and $\mathcal{TR}(V)$ to denote the set of \emph{transition predicates}, i.e., predicates over $\mathcal{T}(V \cup V_{prev})$, where $v_{prev} \in V_{prev}$ iff $v \in V$. Then, we denote by $\preds(V)$ the set of all predicates $\mathcal{ST}(V) \cup \mathcal{TR}(V)$. We also define the set of updates $\mathcal{U}(V)$ of a variable set $V$. Each $U \in \mathcal{U}(V)$ is a function $V \mapsto \mathcal{T}(V)$.

We define the set of valuations over a set of variables $V$ as $\Val(V) = V \mapsto \mathcal{C}$, using $\val \in \Val(V)$ for valuations. For a valuation $\val \in \Val(V)$, we write $\val \models s$, for $s \in \mathcal{ST}(V)$ when $\val$ is a model of $s$. We write $t(\val)$ for $t$ grounded on the valuation $\val$. Given valuations $\val, \val{'} \in \Val(V)$, we write $(\val, \val{'}) \models t$, for $t \in \mathcal{TR}(V)$, when $\val_{prev} \cup \val{'}$ is a model of $t$, where $\val_{prev}(v_{prev}) = \val(v)$ and $dom(\val_{prev}) = V_{prev}$. We say a formula (a Boolean combination of predicates) is satisfiable when there is a valuation that models it. To simplify presentation, we assume $\val \not\models t$ for any $\val$ that does not give values to all the variables of $t$.

\section{Informal Overview}\label{sec:infovr}

\begin{figure}[t]
    \footnotesize
    \begin{minipage}{0.4\linewidth}
    $
    \begin{array}{l}\\
         $V$ = \{\textit{target} : \textit{int} = 0, \textit{floor} : \textit{int} = 0\}\\
         \mathbb{E} = \{\textit{env\_inc}, \textit{door\_open}\}\\
         \mathbb{C} = \{\textit{up}, \textit{down}\}%
    \end{array}
    $
    $
    \begin{array}{l}
        \textbf{Assumptions:}\\
        \text{A1. }G F \textit{door\_open}\\
        \text{A2. }G F \neg \textit{door\_open}
    \end{array}
    $
    $
    \begin{array}{l}
        \textbf{Guarantees:}\\
        \text{G1. } G F \textit{floor} = \textit{target}\\
        \text{G2. } G (\textit{door\_open} \implies (\textit{up} \iff \textit{down}))\\
    \end{array}
    $
    $
    \begin{array}{l}
        \textbf{Objective:}\\
        (A1 \land A2) \implies (G1 \land G2)
    \end{array}
    $
    \end{minipage}%
    \begin{minipage}{0.09\linewidth}
    \hfill
    \end{minipage}%
    \begin{minipage}{0.3\linewidth}
        \scalebox{1}{
    \begin{tikzpicture}[square/.style={}, node distance=5cm]
    \node[state, draw, initial
    ] (s0) {$s_0$};
    \node[state, draw, below = 1cm of s0] (s1) {$s_1$};
    \draw 
    (s0) edge[loop above, left] node{$\begin{array}{l}\\
        \textit{env\_inc} \wedge \textit{door\_open}\\ 
        \mapsto target \inc
    \end{array}\ $} (s0)
    (s0) edge[loop right] node{$
    \begin{array}{l}
        \neg \textit{env\_inc} \wedge \\
        \textit{door\_open}\\ 
        \mapsto target \dec
    \end{array}$} (s0)
    (s0) edge[bend left = 10, right] node{$\neg \textit{door\_open}$} (s1)
    (s1) edge[loop right] node{$\begin{array}{l}
        \textit{up} \wedge \neg \textit{down}\\
        \mapsto floor \inc
    \end{array}$} (s1)
    (s1) edge[loop left] node{$
    \begin{array}{l}
        \textit{down} \wedge \neg\textit{up}\\ \mapsto floor \dec
    \end{array}$} (s1)
    (s1) edge[loop below, right] node{$\ 
    \begin{array}{l}
        \textit{up} \iff \textit{down}\\ \mapsto floor := floor\\
    \end{array}$} (s1)
    (s1) edge[left, bend left = 10] node{$
    \begin{array}{r}
    \textit{door\_open}\ \wedge \\
    \textit{floor} = \textit{target}
    \end{array}$} (s0);
\end{tikzpicture}}

    \end{minipage}
    \caption{Elevator example.}
    \label{fig:example}
\end{figure}

We give a simple instructive LIA example (Fig.~\ref{fig:example}) to illustrate our approach.
Despite its simplicity, we stress that no other existing approach can solve it (see Section~\ref{sec:eval}): since the environment can delay progress by the controller, the resulting objectives are too rich to be expressed by deterministic B\"uchi automata.

On the right is an automaton representing a partial design for an elevator, our arena (see Section~\ref{sec:setting}). A transition labelled $g \mapsto U$ is taken when the guard $g$ holds and it performs the update $U$. Unmentioned variables maintain their previous value.
On the left, we identify input ($\mathbb{E}$) and output ($\mathbb{C}$) Boolean variables.
When guards include these variables, the environment and controller's moves can affect which transitions are possible and which one is taken.
The updates determine how to change the values of other variables ($\mathbb{V}$), which could range over infinite domains.
Thus, the updates of the variables in $\mathbb{V}$ are determined by the interaction between the environment and the controller. 
The desired controller must have a strategy such that, for every possible choice of inputs, it will set the output variables so that the resulting computation satisfies a given LTL objective, encoded on the left as $(\bigwedge_i A_i) \implies (\bigwedge_j G_j)$.
LTL formulas can include quantifier-free first-order formulas over infinite-domain variables (e.g., $floor=target$). 
Notice that this objective includes environment fairness, making this synthesis problem impossible to encode as a deterministic B\"uchi game. 

In our elevator, at state $s_0$ the environment can set a target by controlling variables in $\env$ to increase or decrease \textit{target}. Once a target is set, the environment closes the elevator door (\textit{door\_open}), and the arena transitions to $s_1$. At $s_1$, the system can force the elevator to go up or down one floor, or remain at the same floor. 
This is not a useful elevator: it may never reach the target floor, and it may move with the door open. We desire to control it so that the target is reached infinitely often (G1), and the latter never occurs (G2). 
We also assume aspects of the elevator not in our control to behave as expected, i.e., that     the door is not broken, and thus it opens and closes infinitely often (A1--2).

\smallskip \noindent
\textit{Predicate Abstraction (Defn.~\ref{def:abs})} First, we soundly abstract the arena $\arena$ in terms of the predicates in the specification $(A_1 \wedge A_2) \mathord{\implies} (G_1 \wedge G_2)$, and the predicates, and Boolean variables of the arena
(here, the states in the automaton). That is,\footnote{LIA predicates are normalised to a form using only $\leq$; other relations are macros.} $Pr= \{\textit{floor} \leq \textit{target}, \textit{target} \leq \textit{floor}, s_0, s_1\}$. This abstraction considers all possible combinations of input and output variables and $Pr$, and gives a set of possible predicates holding in the next state (according to the corresponding updates). For example, consider the propositional state $p = s_1 \wedge \mathit{up} \wedge \neg \mathit{down} \wedge \textit{floor} < \textit{target}$. In the automaton, this activates the transition that increments \textit{floor}. Then, satisfiability checking tells us that the successor state is either $p'_1 := s_1 \wedge \textit{floor} = \textit{target}$ or $p'_2 := s_1 \wedge \textit{floor} < \textit{target}$. 

We encode the arena abstraction as an LTL formula $\alpha(\arena, Pr)$ of the form $\mathit{init} \wedge G(\bigvee_{a \in \mathit{abtrans}} a)$, where $\mathit{abtrans}$ is a set of abstract transitions (e.g., $p \wedge X p'_1$ and $p \wedge X p'_2$ are in $\mathit{abtrans}$), and $\mathit{init}$ is the initial state, i.e., $s_0 \wedge \mathit{floor} \mathord{=} \mathit{target}$.

\smallskip \noindent
\textit{Abstract Synthesis.} From this sound abstraction, we create the abstract formula $\alpha(\arena, Pr) \mathord{\implies} \phi$ and treat predicates as fresh \emph{input} Booleans. If this formula were realisable, a controller for it would also work concretely, but it is not: at the abstract state $p$, the environment can always force negation of $\textit{floor} = \textit{target}$.

\smallskip \noindent
\textit{\Counterstrategy Concretisability (Defn.~\ref{defn:compatibility1}).} For an unrealisable abstract problem we will find an abstract \counterstrategy $\moore$. To check whether it is spurious, we model-check if $\arena$ composed with $\moore$ violates the invariant that the predicate guesses of $\moore$ are correct in the arena. Here, $\moore$ admits a finite counterexample $\mathit{ce}$ where the environment initially increments \textit{target}, then moves to $s_1$, and the controller increments \textit{floor}, but $\moore$ wrongly maintains $\textit{floor} < \textit{target}$.

\smallskip \noindent
\textit{Safety Refinement (Section~\ref{ssec:safety}).} By applying interpolation~\cite{DBLP:conf/cav/McMillan06} on  $\mathit{ce}$ we discover new predicates, e.g., $\textit{target} - \textit{floor} \leq 1$, by which we refine the abstraction to exclude $\mathit{ce}$. If we were to continue using safety refinement, we would be attempting to enumerate the whole space, which causes a state-space explosion, given the exponential complexity of predicate abstraction, and the doubly exponential complexity of synthesis.

\smallskip \noindent
\textit{Efficient Encoding (Section~\ref{sec:efficient}).} We manage state-space explosion through a binary encoding of predicates. Note each predicate on a term corresponds to an interval on the reals. For the term $t = \textit{floor} - \textit{target}$, $\textit{floor} \leq \textit{target}$ represents $t \in (-\infty, 0]$. $\textit{target} \leq \textit{floor}$ represents $t \in [0, \infty)$, and $\textit{floor} - \textit{target} \leq 1$ represents $t \in (-\infty, 1]$. These may overlap, but instead we can define formulas whose intervals partition the line $\mathbb{R}$. Here, we get formulas for each interval: $(-\infty, -1], (-1,0], (0,1], (1, \infty)$.
Binary-encoding these reduces the complexity of abstraction and synthesis by an exponential, w.r.t. arithmetic predicates.

\smallskip \noindent
\textit{Liveness Refinements (Section~\ref{ssec:liveness}).} Enumeration is not enough here, given the infinite domain of the variables. Liveness refinements are necessary. Note, once $\moore$ guesses that $\mathit{floor} < \mathit{target}$, it remains in states where $\mathit{floor} < \mathit{target}$ is true. Essentially, we discover a $ce$ in which $\moore$ exercises the loop \textsf{while(floor < target) floor := floor + 1}, and the environment believes it is non-terminating. 
Using known methods to determine the loop is terminating, we construct a monitor for the loop in the abstraction, with extra variables and assumptions. Then a strong fairness constraint that forces the abstraction to eventually exit the loop monitor captures its termination. We term this \textit{structural loop refinement}. 
Note that this is not tied to a specific region in the arena. This allows us to encode more sophisticated loops, beyond what current tools for LTL objectives can do.

With a new synthesis attempt on the refined abstraction, a fresh terminating loop is learned, \textsf{while (target < floor) floor := floor - 1}. Refining accordingly allows us to find a controller and thus solve the problem on the next attempt.

\smallskip \noindent
\textit{Acceleration (Section~\ref{sec:efficient}).} 
The described partitions of the values of a term have a natural well-founded ordering which we can exploit to identify that the controller can force the abstraction to move left or right across the intervals. Consider that if the term $t$ is currently in the interval $(1, \infty)$, and the controller can force strict decrements of $t$, then the value of the $t$ must necessarily eventually move to an interval to the left (unless we have reached the left-most interval). Thus, strict decrements force the value of $t$ to move towards the left of the partition, while strict increments force move towards the right of the partition. Only when the environment can match these increments (decrements) with corresponding decrements (increments) then can this behaviour be prevented.

By adding LTL fairness constraints to represent the described behaviour we can immediately identify a controller, with no further refinements needed.

\begin{toappendix}
Here we give some more detail of the structural refinement for the example in the informal overview.
 
For the loop \textsf{while(floor < target)\{floor := floor + 1\}} we get the following structural loop refinement:
\begin{align*}
    &G((\neg\textit{inloop} \wedge iter\_cond) \implies (X(\textit{floor}_{inc1}) \iff X(\textit{inloop}))\\
    &G((\textit{inloop} \wedge iter\_cond) \implies X((\textit{floor}_{inc1} \implies \textit{inloop}) \wedge (st \implies \textit{inloop})))\\
    &G((\textit{inloop} \wedge \neg iter\_cond) \implies X(\neg \textit{inloop}))\\
    &GF (\neg\textit{inloop}) \vee FG (st \wedge \textit{inloop})
\end{align*}

where $\textit{floor}_{inc1} \mydef \textit{floor} = \textit{floor}_{prev} + 1 \wedge \textit{target} = \textit{floor}_{prev}$, $st \mydef \textit{floor} = \textit{floor}_{prev} \wedge \textit{target} = \textit{floor}_{prev}$, and $iter\_cond \mydef floor < target$. We use one loop variable $\textit{inloop}$ since there is only one step in the loop.

For the loop \textsf{while(target < floor) floor := floor - 1} we get the following structural loop refinement:
\begin{align*}
    &G((\neg\textit{inloop}' \wedge iter\_cond') \implies (X(\textit{floor}_{dec1}) \iff X(\textit{inloop}'))\\
    &G((\textit{inloop}' \wedge iter\_cond') \implies X((\textit{floor}_{dec1} \implies \textit{inloop}') \wedge (st \implies \textit{inloop}')))\\
    &G((\textit{inloop}' \wedge \neg iter\_cond) \implies X(\neg \textit{inloop}'))\\
    &GF (\neg\textit{inloop}') \vee FG (st \wedge \textit{inloop}')
\end{align*}

where $\textit{floor}_{dec1} \mydef \textit{floor} = \textit{floor}_{prev} - 1 \wedge \textit{target} = \textit{floor}_{prev}$, $st \mydef \textit{target} = \textit{floor}_{prev} \wedge \textit{target} = \textit{floor}_{prev}$, and $iter\_cond' \mydef \textit{target} < \textit{floor}$. We use one loop variable $\textit{inloop}'$ since there is only one step in the loop.

\end{toappendix}

\section{Synthesis Setting}\label{sec:setting}

One of our contributions is our special setting that combines arenas and \ltl objectives, unlike existing \ltl approaches which start immediately from \ltl-modulo-theories formulas~\cite{MaderbacherBloem22,10.1145/3519939.3523429,10.1007/978-3-030-25540-4_35}.
We assume a theory, with an associated set of predicates $\preds(V)$ and updates $\mathcal{U}(V)$ over a set of variables $V$. We also assume two disjoint sets of Boolean inputs and outputs $\env$ and $\con$, respectively controlled by the environment and the controller. Then our specifications are \ltl formulas over these variables, $\phi \in \ltl(\env \cup \con \cup Pr_\phi)$, where $Pr_\phi \subseteq \preds(V)$. \ltl formulas talk about an \emph{arena} whose state is captured by the value of $V$, and which modifies its state depending on environment and controller behaviour. Arenas are deterministic; we model (demonic) non-determinism with additional environment variables.
This allows us to encode concretisability checking as invariant checking, rather than the significantly more complex CTL$^*$ model checking.

\begin{definition}[Arena]
    An arena $\arena$ over $V$ is a tuple $\langle V, \val_0, \delta \rangle$, where $V$ is a finite set of variables,
    $\val_0 \in \Val(V)$ is the initial valuation, and 
    $\delta : \mathbb{B}(\mathbb{E} \cup \mathbb{C} \cup \preds(V)) \pfun \mathcal{U}(V)$ is a partial function with finite domain, such that for all $\val \in \Val(V)$ and for every $E\subseteq \mathbb{E}$ and $C\subseteq\mathbb{C}$ there is always a single $f \in dom(\delta)$ such that~$(\val, E \cup C)\models f$.
    An arena is finite when every $v\in V$ is finite.
\end{definition}

Notice that due to the finite domain of $\delta$, an arena $A$ defines a \emph{finite} set of predicates $Pr\subseteq \preds(V)$ and a \emph{finite} set of updates $U\subseteq\mathcal{U}(V)$ that appear in $\delta$. We use the sets $Pr$ and $U$ when clear from the context. 

An infinite concrete word $w \in (\Val(V) \times 2^{\mathbb{E} \cup \mathbb{C}})^\omega$ is a \emph{model} of $\arena$ iff $w(0) = (\val_0, E \cup C)$ (for some $E$ and $C$), and for every $i\geq 0$, $w(i) = (\val_i, E_i \cup C_i)$, then for the unique $f_i\in dom(\delta)$ such that~$(\val_i,E_i\cup C_i)\models f_i$ we have $\val_{i+1}=(\delta(f_i))(\val_i)$. 
We write $L(A)$ for the set of all models of $\arena$.

During our workflow, the words of our abstract synthesis problem may have a different domain than those of the arena. We define these as \textit{abstract words}, and identify when they are concretisable in the arena.
Then, we can define the meaning of (un)realisability modulo an arena in terms of concretisability.

\begin{definition}[Abstract Words and Concretisability]\label{defn:absconcwords}
    For a finite set of predicates $\preds \subseteq \preds(V)$, and a set of Boolean variables $\mathbb{E}'$, such that $\mathbb{E} \subseteq \mathbb{E}'$,
    an \emph{abstract word} $a$ is a word over $2^{\mathbb{E}' \cup \mathbb{C} \cup \preds}$. 
    Abstract word $a$ \emph{abstracts} concrete word $w$, with letters from $\Val(V) \times 2^{\env \cup \mathbb{C}}$, when for every $i$, if $a(i) = E_i \cup C_i \cup \preds_i$, then $w(i) = (\val_i, (E_i \cap \env) \cup C_i)$ for some $\preds_i \subseteq \preds$, $\val_0 \models \cbigwedge_{\preds} \preds_0$, and for $i > 0$ then $(\val_{i - 1}, \val_i) \models \cbigwedge_{\preds} \preds_i$.
    We write $\gamma(a)$ for the set of concrete words that $a$ abstracts.
    We say abstract word $a$ is \emph{concretisable} in an arena $\arena$ when $L(A)\cap \gamma(a)$ is non-empty.
\end{definition}

\begin{definition}[Realisability modulo an Arena]
    A formula $\phi$ in $\ltl(\mathbb{E} \cup \mathbb{C} \cup \preds_\phi)$ is said to be \emph{realisable} modulo an arena $\arena$, when there is a \emph{controller} as a Mealy Machine $\mealy$ with input $\Sigma_{in}=2^{\mathbb{E} \cup \preds_\phi}$ and output $\Sigma_{out}=2^\mathbb{C}$ such that every abstract trace $t$ of \mealy that is concretisable in $\arena$ also satisfies $\phi$.

     A \emph{\counterstrategy} to the realisability of $\phi$ modulo an arena $\arena$ is a Moore Machine \moore with output  $\Sigma_{out}=2^{\mathbb{E} \cup \preds_\phi}$ and input $\Sigma_{in}=2^\mathbb{C}$ such that every abstract trace $t$ of \moore is concretisable in $\arena$ and violates $\phi$.
\end{definition}

\section{Abstract to Concrete Synthesis}\label{sec:abstoconc}

We attack the presented synthesis problem through an abstraction-refinement loop.
We soundly abstract the arena as an \ltl formula that may include fresh predicates and inputs. 
We fix the set of predicates that appear in the objective $\phi$ as $\preds_\phi$,
and the set of predicates and inputs in the abstraction, respectively, as $\preds$ and $\mathbb{E}'$, always such that $\preds_\phi \subseteq \preds$ and $\mathbb{E} \subseteq \mathbb{E}'$.

\begin{definition}[Abstraction]\label{def:abs}
    Formula $\alpha(A, \preds)$ in $\ltl(\mathbb{E}' \cup \mathbb{C} \cup \preds)$ abstracts arena $\arena$ if for every $w \in L(\arena)$ there is $a \in L(\alpha(\arena, \preds))$ such that $w \in \gamma(a)$. 
\end{definition}

$\alpha(\arena, \preds)$ is a standard predicate abstraction~\cite{GS97cas}. Given the lack of novelty, we refer to Appendix~\ref{sec:predabs} for the full details.
Note, $\alpha(\arena, \preds)$ can be non-deterministic, unlike $\arena$. 
Constructing it is essentially an ALLSAT problem: given a transition, we identify sets from $2^{\preds}$ that can be true before the transition and, for each of these, sets of $2^{\preds}$ that can hold after the transition. However, we construct these sets incrementally, adding predicates as we discover them; and improve on the space/time complexity with a binary encoding (Section~\ref{sec:efficient}).

Given abstraction $\alpha(\arena, \preds)$, we construct a corresponding sound \ltl synthesis problem, $\alpha(\arena, \preds) \implies \phi$, giving the environment control of the predicates in $\alpha(\arena, \preds)$. We get three possible outcomes from attempting synthesis of this: (1) it is realisable, and thus the concrete problem is realisable; (2) it is unrealisable and the \counterstrategy is concretisable; or (3) the \counterstrategy is not concretisable. We prove theorems and technical machinery essential to allow us to determine realisability (1) and unrealisability (2). In case (3) we refine the abstraction to make the \counterstrategy unviable in the new abstract problem.

\begin{theoremrep}[Reduction to \ltl Realisability]
    For $\phi$ in $\ltl(\mathbb{E}\cup \mathbb{C}\cup \preds_\phi)$ and an abstraction $\alpha(\arena, \preds)$ of $\arena$ in $\ltl(\mathbb{E}'\cup \mathbb{C} \cup \preds)$, if $\alpha(\arena, \preds) \implies \phi$ is realisable over inputs $\mathbb{E}' \cup \preds$ and outputs $\mathbb{C}$, then $\phi$ is realisable modulo $\arena$.
\end{theoremrep}
\begin{proof}
    This follows immediately from the soundness of the abstraction.
\end{proof}

However, an abstract \counterstrategy $\moore$ may contain unconcretisable traces, since abstractions are sound but not complete. To analyse $\moore$ for concretisability, we define a simulation relation between states of the concrete arena and states of $\moore$, capturing whether each word of $\moore$ is concretisable. 
Recall, a set of predicates $\preds$ is the union of a set of state predicates, $ST$ (describing one state), and transition predicates, $TR$ (relating two states), which require different treatment.

\todo{S: reviewer 4 wants us to make this definition more readable}
\begin{definition}[\Counterstrategy Concretisability]\label{defn:compatibility1}
	Consider a \counterstrategy as a Moore Machine $\moore=\pair{S,s_0,{\prop}_{in},{\prop}_{out},\rightarrow,out}$, and an arena $\arena$,
	where ${\prop}_{in}=2^{\mathbb{C}}$ and ${\prop}_{out}=2^{\mathbb{E}' \cup \preds}$.

	\noindent
	\emph{Concretisability} is defined through the simulation
	relation ${\preceq_A} \subseteq \Val \times S$: 

    \noindent
    For every valuation $\val$ that is simulated by a state $s$, $\val \preceq_{A} s$, where $out(s) = E \cup ST \cup TR$, it
	holds that: 
    \begin{enumerate}
    \item the valuation satisfies the state predicates of $s$: $\val \models \cbigwedge ST$, and 
    \item for every possible controller output $C\subseteq \mathbb{C}$: let $\val_C = \delta(\val, (E \cap \env) \cup C)$, $s_C$ be s.t. $s \xrightarrow{C} s_C$, and $TR_{C}$ be the transition predicates in $out(s_C)$, then 
    \begin{enumerate}
        \item the transition predicates of $s_C$ are satisfied by the transition $(\val, \val_C) \models \cbigwedge TR_{C}$, and
        \item the valuation after the transition simulates the $\moore$ state after the transition: $\val_C \preceq_{A} s_C$.
    \end{enumerate}
     \end{enumerate}

\smallskip
\noindent
\moore is \emph{concretisable} w.r.t. $\arena$ when $\val_0 \preceq_{A} s_0$, for $\arena$'s initial valuation $\val_0$. 
\end{definition}

\begin{toappendix}
\begin{lemmarep}\label{lem:csconcfromwordconc}
    An abstract counterstrategy \moore is concretisable w.r.t. $\arena$ iff every trace $t \in L(\moore)$ is concretisable w.r.t. $\arena$.
\end{lemmarep}
\begin{proof}
    This follows from Defn.~\ref{defn:absconcwords}.
\end{proof}
\end{toappendix}

With concretisability defined, we then have a method to verify whether an abstract \counterstrategy is also a concrete \counterstrategy.

\begin{theoremrep}[Reduction to \ltl Unrealisability]
    Given arena abstraction $\alpha(\arena,\preds)$, if $\alpha(\arena,\preds) \implies \phi$ is unrealisable with a \counterstrategy \moore and \moore is concretisable w.r.t. $\arena$, then $\phi$ is unrealisable modulo $\arena$.
\end{theoremrep}
\begin{proof}
    This follows immediately from Defns.~\ref{defn:absconcwords} and \ref{defn:compatibility1}.
\end{proof}

In practice, we encode \counterstrategy concretisability as a model checking problem on the composition of the \counterstrategy and the arena, with the required invariant that predicate values chosen by the \counterstrategy hold on the arena. 
Conveniently, this also gives witnesses of unconcretisability as finite counterexamples (rather than infinite traces), which we use as the basis for refinement. Crucially, this depends on the choices of the environment/controller being finite, which also gives us semi-decidability of finding non-concretisability.
\begin{propositionrep}
    \Counterstrategy concretisability is encodable as invariant checking, and terminates for finite problems and non-concretisable \counterstrategies.
\end{propositionrep}
\begin{proof}
    Given an arena $\arena$, a formula $\phi$, and a \counterstrategy $CS$ with predicates $\preds$, we compose $\arena$ with $CS$, giving a program $A \times CS$, in the following manner: 
    the variables of $A \times CS$ are the variables of $\arena$, the set of states of $CS$, and fresh Boolean variables for each predicate in $\preds$, i.e., $v_p$ for each predicate $p \in \preds$ (we denote this set by $V_{\preds}$); initially all the $CS$ state variables are false except the initial state $s_0$; and all predicate variables expected to be true by $s_0$ are set to true, and all others to false.
    
    For each transition $g \mapsto U$ in the arena and for each transition $s_i \xrightarrow{C} s_{i+1}$ in $CS$, such that $out(s_i) = (E_i, Pr_i)$, and $out(s_i) = (E_{i+1}, Pr_{i+1})$, there is a transition $(g \wedge (\cbigwedge_{S} \{s\}) \wedge (\cbigwedge E_{i+1}) \wedge (\cbigwedge_{V_{Pr}} V_{Pr_i})) \mapsto U'$ in $A \times CS$. Update $U'$ consists of $U$, extended with the following updates: $\{s_{i+1} := true;\}$, $\{s_j := \textit{false} \mid j \neq i + 1\}$, $\{v_{p} := true \mid p \in Pr_{i+1}\}$, $\{v_{p} := false \mid p \not\in Pr_{i+1}\}$, and $\{v_{prev} := v \mid v \in V\}$. The transitions of $A \times CS$ are exactly these transitions.

    We can see that this program satisfies the invariant $G \bigwedge (v_p \iff p)$ iff $CS$ is concretisable on $\arena$. Moreover, for finite arenas this program is finite, for which model checking is decidable. 
    
    Assume the \counterstrategy is not concretisable, then by Defn.~\ref{defn:compatibility1} there must be a finite counterexample. Moreover, the arena has one initial state, and only allows for finite branching in each time step. Thus, if there is a counterexample, it will be found in finite time. \qed
\end{proof}

\begin{propositionrep}\label{prop:cex}
    A non concretisable \counterstrategy induces a finite counterexample $a_0, \ldots, a_k \in (2^{\env \cup \con \cup \preds})^*$ and concretisability fails locally only on $a_k$.
\end{propositionrep} 
\begin{proof}
    This follows easily from the fact that concretisability checking can be encoded as invariant checking.
\end{proof}

\begin{algorithm}[t]
	\caption{Synthesis algorithm based on abstraction refinement.\label{alg:capsafety}}
    \scriptsize
	\DontPrintSemicolon
	\SetKwFunction{FMain}{synthesise}
	\SetKwProg{Fn}{Function}{:}{}
	\Fn{\FMain{$\arena$, $\phi$}}{
		$\preds, \psi := \preds_\phi, \textit{true}$\;\label{alg:setup}
	\While{\true}{
      $\phi^\arena_\alpha := (\alpha(\arena, \preds) \wedge \psi) \implies \phi$\;\label{alg:abstract}
      \lIf{$\mathtt{realisable}(\phi^\arena_\alpha,\mathbb{E} \cup \preds,\mathbb{C})$}{
				\Return $(\true,
				\texttt{strategy}(\phi^\arena_\alpha,\mathbb{E} \cup \preds,\mathbb{C}))$\label{alg:return-real}
	   }
      $\Cs:=\texttt{counter\_strategy}(\phi^\arena_\alpha,\mathbb{E} \cup \preds,\mathbb{C})$\;
		
      \lIf{$\mathtt{concretisable}(\phi, \arena,\Cs)$}{
					\Return $(\false, \Cs)$\label{alg:return-unreal}
				}
        $\preds', \psi' := \texttt{refinement}(\arena,\Cs)$\;\label{alg:refine}
        $\preds, \psi := \preds \cup \preds', \psi \wedge \psi'$\label{alg:extend}
		}
	}
\end{algorithm} 
\noindent\textit{Synthesis Semi-Algorithm.} Alg.~\ref{alg:capsafety} shows our high-level approach.
Taking an arena $\arena$ and an LTL formula $\phi$, it maintains a set of predicates $\preds$ and an \ltl formula $\psi$. When the abstract problem (in terms of $\preds$) is realisable, a controller is returned (line~\ref{alg:return-real}); otherwise, if the \counterstrategy is concretisable, it is returned (line~\ref{alg:return-unreal}). If the \counterstrategy is not concretisable, we refine the abstraction to exclude it (line~\ref{alg:refine}), and extend $\preds$ with the learned predicates, and $\psi$ with the new \ltl constraints (line~\ref{alg:extend}). 
Alg.~\ref{alg:capsafety} diverges unless it finds a (counter)strategy.

\begin{toappendix}
\subsection{Predicate Abstraction}\label{sec:predabs}

We define an abstraction of the arena in terms of a set of predicates $\preds$. Initially, $\preds$ is exactly the set of predicates appearing in the desired formula $\phi$.\footnote{A technical detail is that Boolean variables in $V$ are also in $\preds$.}

The arena abstraction then focuses on abstracting the symbolic transition relation $\delta_{sym}$ of the arena in terms of $\preds$, such that every symbolic transition has corresponding abstract transitions. We rely on satisfiability checking to compute this abstraction. Moreover, given that we have an initial variable valuation, we give a sound and complete abstraction for the initial transition. This will be crucial later to ensure progress of safety refinement.

\begin{definition}[Abstracting the Initial Transition]\label{def:initabs}
    Given a set of predicates $\preds$ and a arena $\arena$, the \emph{initial transition abstraction} of $\arena$ w.r.t. $\preds$ is the relation 
    $\iota_{\preds} \subseteq 2^{\mathbb{E} \cup \mathbb{C}} \times  2^{\preds}$, such that $(E\cup C, Pr_{E,C}) \in \iota_{\preds}$ iff there exists $f\in dom(\delta_{sym})$ such that $(\cbigwedge (E\cup C)) \wedge val_0 \wedge f$ is true and if $U=\delta(f)$ then $(\val_0, U(\val_0)) \models (\cbigwedge_{\preds}Pr_{E,C})$.
\end{definition}

Notice that $(E\cup C,Pr_{E,C})\in \iota_{\preds}$ iff $\delta(\val_0,E \cup C)\models (\cbigwedge_{\preds}Pr_{E,C})$. Furthermore, due to determinism of $\arena$, for every $E\cup C$ there is a unique $Pr_{E,C}$ such that $(E\cup C,Pr_{E,C})\in \iota_{\preds}$.

\begin{definition}[Abstracting Transitions]\label{def:abstrans}
    Given a set of predicates $\preds$ and an arena $\arena$, the \emph{abstract transition} of $\arena$ w.r.t. $\preds$ is a relation $\delta_{\preds} \subseteq 2^{\mathbb{E} \cup \mathbb{C} \cup \preds} \times {2^{\preds}}$, such that $(E \cup C \cup Pr^0, Pr^1) \in \delta_{\preds}$ iff there exists $f \in dom(\delta)$ such that 
    $(\cbigwedge (E \cup C \cup Pr^0)) \wedge f$ is satisfiable and if $U=\delta(f)$ then $(\cbigwedge Pr^0_{prev}) \wedge f_{prev} \wedge (\bigwedge_{v := t \in U} v = t_{prev}) \wedge (\cbigwedge Pr^1))$ is satisfiable as well. 
    
    We further assume this is reduced up to reachability, such that $(E \cup C \cup Pr^0) \in dom(\delta_{\preds})$ iff $Pr^0 \in ran(\delta_{\preds})$ or $Pr^0 \in ran(\iota_{\preds})$.
\end{definition}

Note that $(E\cup C \cup Pr,Pr')\in \delta_{\preds}$ if and only if there exist valuations $\val$ and $\val'$ such that $\val \models (\cbigwedge_{\preds}(Pr))$, $(\val, \val') \models (\cbigwedge_{\preds}Pr')$, and $\delta(\val,E\cup C)=\val'$.

Based on these we define a formula in $LTL(\mathbb{E} \cup \mathbb{C} \cup \preds)$ that abstracts the arena. Let $Pr_\iota$ be the set of predicates true for $\val_0$: $\val_0 \models (\cbigwedge_{\preds}Pr_\iota)$.

\begin{definition}[Safety Abstraction]
    The abstract characteristic safety formula $\alpha(P, \preds)$ w.r.t. a set of predicates $\preds$ is the conjunction of: 
    $\cbigwedge Pr_{\iota}$ characterising the predicates holding initially, 
    $\bigvee_{(E \cup C,Pr)\in \iota_{\preds}} (\cbigwedge E \cup C) \wedge X (\cbigwedge Pr)$ characterising the initial transition, and
    $G\left (\bigvee_{(S,Pr') \in \delta_{\preds}} (\cbigwedge S \wedge X\cbigwedge Pr')\right )$ characterising all the other transitions.
\end{definition}

Note that the initial transitions captured by the second conjunct are a special case of the full transition abstraction in the third conjunct.

\begin{propositionrep}\label{prop:predabscor}
    The formula $\alpha(P, \preds)$ is an abstraction of $\arena$.
\end{propositionrep}
\begin{proof}
    Recall that a formula over $\mathbb{E} \cup \mathbb{C} \cup \preds$ is an abstraction of a arena $\arena$ if for all concrete models of the arena there is an abstract word in the abstraction that abstracts the concrete word.
    
    Consider a word $w_a$ in $L_{\preds}(A)$. We claim there is a word $a$ such that $w_a \in \gamma(a)$ and $a$ is in $\alpha(P, Pr)$. For the initial state it should be clear that the initial condition $\cbigwedge Pr_{\iota}$ ensures the initial arena state is properly abstracted. The rest of the abstraction abstracts transitions, thus we prove its correctness by induction on pairs of successive letters. Throughout, for the concrete word $w_a$, we set $w_a(i) = (\val_i, E_i \cup C_i)$. 
    
    For the base case, we consider the first two letters of $a$, $a(0)$ and $a(1)$. Since $a$ is concretisable it follows easily that $Pr_\iota \subseteq a(0)$, and that there is a transition $f \mapsto U$ such that $f(w_a(0))$, $\val_{1} = U(\val_0)$, and $(\val_0, \val_{1}) \models a(1)$. From this it follows that this initial transition is captured by Defn.~\ref{def:initabs}.

    For the inductive case, consider $a(i)$ and $a(i+1)$. By Defn.~\ref{defn:absconcwords}, for $j \in \{i, i+1\}$ we have that $w_a(j) = (\val_j, E_j \cup C_j)$, then $a(j) = E_j \cup C_j \cup \preds_j$ for some predicate set $\preds_j \subseteq \preds$, and $(\val_i, \val_{i+1}) \models \cbigwedge_{\preds} \preds_{i + 1}$. Concretisability of the word, ensures there is a transition $f \mapsto U$ such that $f$ and $a(i)$ is satisfiable ($w_a(i)$ is a model for this). Consider that $\bigwedge Pr_i \wedge f$ is satisfiable, implying that $\bigwedge (Pr_i)_{prev} \wedge f_{prev}$. Moreover, consider that $\val_{i+1} = U(\val_i)$, implying that  $(\bigwedge (Pr_i)_{prev}) \wedge f_{prev} \wedge (\bigwedge_{v := t \in U} v = t_{prev}) \wedge (\cbigwedge Pr_{i+1})$ is also satisfiable, as required by Defn.~\ref{def:abstrans}.
\end{proof}

The complexity of this construction, ignoring satisfiability checking, is at worst exponential in the size of $\env \cup \con \cup \preds$. Depending on the theory, satisfiability checking may increase this complexity. For LIA, this is exponential, leaving the complexity lower than the complexity of LTL reactive synthesis.

\end{toappendix}

\section{Refinement}\label{sec:ref}

We now present the two refinements on which our iterative approach relies, based on an analysis of a discovered \counterstrategy.
These refinements soundly refine the abstraction with predicates and/or new \ltl constraints such that
similar counterexamples will not be re-encountered in the next iteration.\footnote{We prove a progress theorem for each refinement in Appendix~\ref{apx:abs}.}

\subsection{Safety Refinement}\label{ssec:safety}

Consider a \counterstrategy \moore and a counterexample $ce = a_0,a_1,\ldots, a_k$.
The transition from $a_{k-1}$ to $a_k$ induces a mismatch between the concrete arena state and \moore's desired predicate state.
It is well known that interpolation can determine sufficient state predicates to make \moore non-viable in the fresh abstract problem; we give a brief description for the reader's convenience.
Let $p_i = \cbigwedge_{\preds} ({a_i \cap \preds})$, with each variable $v$ replaced by a fresh variable $v_i$, and each variable $v_{prev}$ by $v_{i-1}$. Similarly, let $g_i$ and $u_i$ be respectively the corresponding symbolic transition guard and update (i.e., $\delta(g_i)=u_i$), such that all updates $v := t$ are rewritten as $v_{i+1} = t_i$, where term $t_i$ corresponds to $t$ with every variable $v$ replaced by $v_i$. 

In order to characterize the mismatch between the arena and its abstraction, we construct the following formulas.
Let $f_0 = \val_0 \wedge p_0 \wedge g_0 \wedge u_0$, where we abuse notation and refer to $\val_0$ as a Boolean formula.
For $1 \leq  i < k$, let $f_i = p_i \wedge g_i \wedge u_i$, while $f_k = p_k$. 
Then $\bigwedge_{i=0}^k f_i$ is unsatisfiable.
Following McMillan~\cite{DBLP:conf/cav/McMillan06}, we construct the corresponding set of \emph{sequence interpolants} $I_0,...,I_{k-1}$, where $f_0 \implies I_1$, $\forall 1 \leq i < k . I_{i} \wedge f_i \implies I_{i+1}$, $I_{k-1} \wedge f_k$ is unsatisfiable, 
as all the variables of $I_i$ are shared by both $f_{i-1}$ and $f_i$. 
From these we obtain a set 
of state predicates $I(ce)$ by removing the introduced indices in each $I_i$. Adding $I(ce)$ to the abstraction refines it to make the \counterstrategy unviable.

\begin{toappendix}
\label{apx:abs}
\begin{theorem}[Existence of Sequence Interpolants~\cite{DBLP:conf/cav/McMillan06}]\label{thm:seqint}
    For a sequence of formulas $f_0,\ldots, f_k$, such that $\bigwedge_{i=0}^k f_i$ is unsatisfiable and for every $i,j$ either $|i-j|\leq 1$ or $f_i$ and $f_j$ do not share variables, then there is a set of \textit{sequence interpolants} $I_0,\ldots, I_{k-1}$, where $f_0 \Longrightarrow I_0$, $\forall 1\leq i < k . f_i \wedge I_{i} \Longrightarrow I_{i+1}$, and $I_{k - 1} \wedge f_k$ is unsatisfiable. Furthermore, the variables of each $I_i$ appear in both $f_{i}$ and $f_{i+1}$.
\end{theorem}
\end{toappendix}

\begin{toappendix}
\begin{propositionrep}[Safety Refinement Progress]\label{prop:safeprogress}
    For an abstraction $\alpha(\arena, Pr)$ that allows a \counterstrategy $\moore$ with a finite counterexample $ce$, then $\alpha(\arena, Pr \cup I(ce))$ does not allow counterexamples that induce the same refinement.
\end{propositionrep}
\begin{proof}
    Suppose the abstraction $\alpha(\arena, Pr \cup I(ce))$ contains an unconcretisable w.r.t. $\arena$ word $a$ with a finite prefix $a_0,\ldots, a_k$, such that concretisability fails only on $a_k$. Suppose further that interpolants $I^a = I(a_0,\ldots, a_k)$ corresponding to this trace are equal (or a subset) of $I(ce)$. 
    
    Note how the initial transition abstraction ensures $I^a_0$ is always true in the first step, thus $a_0$ must guess $I^a_0$ to be true. Similarly, by Thm.~\ref{thm:seqint} and Defn.~\ref{def:abstrans} the correct guesses of interpolants must be maintained throughout. Then, $a_k$ must guess $I^a_k$ to be true, however this creates a contradiction, since Defn.~\ref{def:abstrans} requires the predicate guesses to be satisfiable, but Thm.~\ref{thm:seqint} ensures $I^a_k$ is not satisfiable with $a_k$. \qed
\end{proof}
    
\end{toappendix}

\begin{toappendix}
    
\begin{theoremrep}
    Alg.~\ref{alg:capsafety} with safety refinement terminates on finite arenas.
\end{theoremrep}
\begin{proof}
    Note that a finite arena $P$ has a finite number of possible variable valuations, and thus $\delta_P$ is finite. Then model checking is decidable. Finding sequence interpolants is also decidable.
    
    Moreover, recall that given a counterexample, the interpolants $I$ learned through safety refinement always strictly refine the abstraction, Prop.~\ref{prop:safeprogress}. Consider a predicate set $\preds$ and $\preds' = \preds \cup I$, where $I$ is a set of interpolants discovered through analysing a counterexample. Consider also that $\delta_{\preds}$ is an abstraction for $\delta$, such that each element $(E \cup C \cup Pr, Pr') \in \delta_{\preds}$ has a corresponding concrete finite set of transitions in $\delta$. Upon adding $I$ to the abstraction, each original abstract transition is replicated for each subset of $I$. Each of these new abstract transitions partition a subset of the original set of concrete transitions of the corresponding abstract transition between them.
    
    Then, $\delta_{Pr}$ accepts strictly more valuation pairs $((E \cup C, \val), \val')$ than $\delta_{Pr'}$. Given there is a finite set of such valuation pairs, and refinement always makes progress, then refinement can only be repeated for a finite amount of steps. This ensures there cannot be an infinite chain of discovered spurious counterstrategies, and thus a concretisable counterstrategy or controller are eventually found. \qed
\end{proof}
\end{toappendix}

\subsection{Liveness Refinement}\label{ssec:liveness}
Relying solely on safety refinement
results in non-termination for interesting problems (e.g., Fig.~\ref{fig:example}).
To overcome this limitation,
we propose \textit{liveness refinement}. 
Our main insight is that if the counterexample exposes a spurious lasso in 
the \counterstrategy, then we can encode its termination as a liveness property.

\smallskip \noindent
\emph{Lassos and Loops.} A counterexample $ce = a_0,\ldots, a_k$ \emph{induces} a lasso in \moore when 
\begin{wrapfigure}{r}{2.9cm}
\vspace{-10mm}
\centering
\hspace*{0.1cm}
\begin{footnotesize}
\begin{alltt}
  V = *
  \textbf{assume} \(\val\sb{j}\)
  \textbf{while} \(\cbigwedge(a_j\cap\preds)\)
     \textbf{assume} \(g\sb{j}\)
     V = \(U\sb{j}\)(V)
     \ldots
     \textbf{assume} \(g\sb{k-1}\)
     V = \(U\sb{k-1}\)(V)
    \end{alltt}
\end{footnotesize}
\vspace{-0.7cm}
\caption{$ce$ loop.\label{alg:loop}}
\vspace*{-8mm}
\end{wrapfigure}%
it %
corresponds to a path $s_0,\ldots, s_{k}$ in \moore, where $s_{k} = s_j$ for some $0 \leq j < k$. 
We focus on the last such $j$. Here, for simplicity, we require that concretisation failed due to a wrong state predicate guess.
We split the counterexample into two
parts: 
a stem $a_0,\ldots, a_{j-1}$, and a loop $a_{j},\ldots, a_{k-1}$. 
Let $g_j\!\mapsto U_j,\ldots, g_{k-1} \!\mapsto U_{k-1}$ be the corresponding applications of $\delta$ and let $\val_j$ be the arena state at step $j$.

The counterexample proves that the while-program in Fig.~\ref{alg:loop} terminates (in one iteration). To strengthen the refinement, we try to weaken the loop (e.g., expand the precondition) such that it still accepts the loop part of $ce$ while terminating. We formalise loops to be able to formalise this weakening.

\begin{definition}[Loops]
    A \emph{loop} is a tuple $l = \langle V, \textit{pre}, \textit{iter\_cond}, \textit{body}\rangle$, where $\textit{pre}$ and $\textit{iter\_cond}$ are Boolean combinations of predicates over variables $V$, and $\textit{body}$ is a finite sequence of pairs $(g_i, U_i)$, where $g_i \in \preds(V)$ and $U_i \in \mathcal{U}(V)$.

     A finite/infinite sequence of valuations $\vals = \val_0,\val_1,\ldots$ is an execution of $l$, $\vals \in L(l)$, iff $\val_0 \models pre$, for all $i$ such that $0 \leq i < |\vals|$, where $n = |\textit{body}|$, then $\val_{i} \models g_{i \bmod n}$, $\val_{i+1} = U_{i \bmod n}(\val_i)$ and if $i \bmod n=0$ then $\val_i \models \textit{iter\_cond}$. 
     We say a loop is \emph{terminating} if all of its executions are finite.
\end{definition}

\begin{definition}[Weakening]
    Loop $l_1 = \langle V_1, \textit{pre}_1, \textit{ic}_1, \textit{body}_1 \rangle$ \emph{is weaker than} $l_2 = \langle V_2, \textit{pre}_2, \textit{ic}_2, \textit{body}_2 \rangle$ when:
    \begin{inparaenum}
        \item $V_1 \subseteq V_2$;
        \item $\textit{pre}_2 \mathord{\implies} \textit{pre}_1$ and $\textit{ic}_2 \mathord{\implies} \textit{ic}_1$;
        \item $|\textit{body}_1| = |\textit{body}_2|$;
        \item for $w_2 \in L(l_2)$ there is $w_1 \in L(l_1)$ such that $w_2$ and $w_1$ agree on $V_1$.
    \end{inparaenum}
A weakening is \emph{proper} if both $l_1$ and $l_2$ terminate.
\end{definition}

 \noindent
\emph{Heuristics.} We attempt to find loop weakenings heuristically. In all cases we reduce $iter\_cond$ to focus on predicates in $a_k$ that affect concretisability. We also remove variables from the domain of the loop that are not within the cone-of-influence~\cite{ClGP99-Mc} of $iter\_cond$. We then attempt two weaker pre-conditions: (1) \textit{true}; and (2) the predicate state before the loop is entered in the \textit{ce}. 
We check these two loops, in the order above, successively for termination (using an external tool). The first loop proved terminating ($l(ce)$) is used as the basis of the refinements.

\smallskip
\noindent

\smallskip \noindent
\emph{Structural Loop Refinement.} We present a refinement that monitors for execution of the loop and enforces its termination.

We define some predicates useful to our definition. For each transition in the loop we define a formula that captures when it is triggered: $cond_0 \mydef iter\_cond \wedge g_0$ and $cond_i \mydef g_i$ for all other $i$. For each update $U_i$, we define a conjunction of transition predicates that captures when it occurs: recall $U_i$ is of the form $v^0 := t^0,\ldots, v^j := t^j$, then we define $p_i$ as $v^0 = t^0_{prev} \wedge\ldots \wedge v^j = t^j_{prev}$. This sets the value of variable $v^k$ to the value of term $t^k$ in the previous state. We further define a formula that captures the arena stuttering modulo the loop, $st \mydef \bigwedge_{v \in V_l} v = v_{prev}$, where $V_l$ is the set of variables of the loop.
A technical detail is that we require updates in the loop $l(ce)$ to not stutter, i.e., $U(\val) \neq \val$ for all $\val$.
Any loop with stuttering can be reduced to one without, for the kinds of loops we consider. 
Thus, here $p_i\wedge st$ is contradictory, for all $i$.

\begin{definition}[Structural Loop Refinement]
    Let $l$ be a terminating loop,
    and $cond_{i}$, $p_i$, and $st$ (for $0 \leq i < n$) be as defined above. Assume fresh variables corresponding to each step in the loop $inloop_0,\ldots, inloop_{n-1}$, and $inloop = inloop_0 \vee\ldots \vee inloop_{n-1}$. 
    
    The structural loop abstraction $\alpha_{loop}(A, l)$ is the conjunction of the following:
    
    \begin{compactenum}
        \item Initially we are not in the loop, and  we can never be in multiple loop steps at the same time: 
        $\textstyle \neg inloop \wedge \bigwedge_i G(inloop_i \implies \neg \bigvee_{j \neq i} (inloop_j))$;
        \item The loop is entered when $pre$ holds and the first transition is executed: $\textstyle G(\neg inloop \implies (
            (pre \wedge cond_0 \wedge X (p_0)) \iff X(inloop_{1})))$;
        \item At each step, while the step condition holds, the correct update causes the loop to step forward, stuttering leaves it in place, otherwise we exit: 
        \[\textstyle \bigwedge_{0 \leq i < n} G \left ( (inloop_i \wedge cond_i) \implies 
        X\left(\begin{array}{l}
            (p_i \implies inloop_{i + 1 \% n}) \wedge \\
            (st \implies inloop_{i}) \wedge \\
            (\neg(st \vee p_i) \iff \neg inloop)
        \end{array}\right)\right)\textit{;}\]
        \item At each step, if the expected step condition does not hold, we exit: 
        
        $\textstyle \bigwedge_{0 \leq i < n} G((inloop_i \wedge \neg cond_i) \implies X \neg inloop)$; and
        \item The loop always terminates, or stutters: $\textstyle G F (\neg inloop) \vee \bigvee_{i}  F G (st_i \wedge inloop_i)$.
    \end{compactenum}
\end{definition}

Note the fresh propositions ($\textit{inloop}_i$) are controlled by the environment. The \ltl formulas 1--4 monitor for the loop, exiting if a transition not in the loop occurs, and progressing or stuttering in the loop otherwise. \ltl formula 5 enforces that the loop is exited infinitely often, or that the execution stutters in the loop forever. This ensures that the abstract \counterstrategy is no longer viable.

\begin{toappendix}
\begin{propositionrep}[Structural Loop Refinement Correctness]
    For a terminating loop $l$, and a set of predicates $\preds_l$ that consists of exactly all the atomic predicates over arena variables in $pre$ and $\alpha_{loop}(P, l)$, then $\alpha(\arena, Pr \cup \preds_l) \wedge \alpha_{loop}(\arena, l)$ is an abstraction of $\arena$.
\end{propositionrep}
\begin{proof}
    Consider a word $a \in AL_\preds(\arena)$. We know this is in $\alpha(\arena, Pr \cup \preds_l)$ by Prop.~\ref{prop:predabscor}, what is left to show is that it is in $\alpha_{loop}(\arena, l)$, modulo some additions of $inloop_i$ variables. It should be easy to see that conditions 1-6 do not put any restrictions on the variable state space. Only condition 7 has the potential to eliminate arena words unsoundly. 
    
    Let $a$ be a word in $AL_\preds(\arena)$ that has no counterpart in $\alpha_{loop}(\arena, l)$. Then, it must satisfy the negation of condition 7, i.e., $FG(inloop) \wedge \bigwedge_i G F (\neg(st_i \wedge inloop_i))$, so that the word eventually remains in the loop without stuttering. Then $a$ must have a maximal $k$ such that the suffix $a_k$ has that $pre \wedge iter\_cond$ is satisfiable with $a(k)$, and $a(k+1)$ satisfies $p_0$, since this is the only way for $inloop$ to become true (formula 3). Then, at each point in time either $4$ or $5$ hold and keep $inloop$ true. Moreover, at each loop step there must be a finite amount of stuttering (as required by negation of 7). Thus $a_k$ corresponds, up to stuttering, to the loop with precondition $pre$, iteration condition $iter\_cond$, and body $(g_0, U_0),\ldots,(g_n,U_n)$ (given the correspondence of the predicates $p_i$ to these guarded updates). Note that any concretisation of $a_k$ must not exit from this loop. However, by assumption this loop is terminating, creating a contradiction. \qed 
\end{proof}

\begin{propositionrep}[Structural Loop Refinement Progress]
    For an abstraction $\alpha(\arena, Pr)$ allowing a \counterstrategy $\moore$ with a finite counterexample $ce$ that induces a lasso in $\moore$ and a corresponding loop $l$, the abstraction $\alpha(\arena, Pr \cup \preds_l) \wedge \alpha_{loop}(\arena, l)$ does not allow counterexamples that induce the same refinement.
\end{propositionrep}
\begin{proof}
    Suppose that the abstraction $\alpha(\arena, Pr \cup \preds{l}) \wedge \alpha_{loop}(\arena, l)$ contains an unconcretisable w.r.t. $\arena$ word $a$, with a prefix $a_0,\ldots, a_l,\ldots, a_k$, (such that $l < k$) such that concretisability fails due to a state predicate mismatch on $a_k$, and this exercises a lasso in the counterstrategy $s_0,\ldots,s_l,\ldots, s_k$, such that $s_k = s_l$, and the suffix of the word is thus of the form $(a_l,\ldots, a_k-1)^\omega$. Suppose further that $a$ also has the corresponding loop $l$, and requires the same refinement. 

    If $a$ guesses $pre$ wrongly (false) at $a_l$ then concretisability will fail at $a_l$ rather than at $a_k$. Thus we assume $pre$ is guessed correctly, and similarly for the iteration condition at $a_l$. Moreover, all of the transition predicates ($p_i$ and $st_i$) must be guessed correctly, otherwise the mismatch is not a state predicate mismatch. However, then if all these are guessed correctly $a$ is a witness that the abstraction allows words that go through the loop (as captured by conditions 1-6), and remains in the loop, violating condition $7$, violating $\alpha_{loop}(\arena, l)$. \qed
\end{proof}
    
\end{toappendix}

\section{Efficient Encoding and Acceleration}\label{sec:efficient}

The problem we tackle is undecidable, but we rely on decidable sub-routines of varying complexity: predicate abstraction (exponential in the number of predicates) and finite synthesis (doubly exponential in the number of propositions, of which predicates are a subset). Here we present an efficient binary encoding of predicates of similar forms that 
\begin{inparaenum}[(1)]
\item 
reduces the size of and the satisfiability checks needed to compute the abstraction from exponential to polynomial, and 
\item
reduces complexity of abstract synthesis from doubly to singly exponential, 
\end{inparaenum}when restricted to predicates. 
Moreover, this encoding allows us to identify fairness assumptions refining the abstraction, which significantly accelerate synthesis. Computing this encoding only involves simple arithmetic, but we have not encountered previous uses of it in literature.

We collect all the known predicates over the same term, giving a finite set of predicates $P_t = \{t \bowtie c_0, ..., t \bowtie c_n\}$, where $t$ is a term only over variables, $\bowtie \in \{<,\leq\}$ and each $c_i$ is a value. 
W.l.g.~we assume $t\bowtie c_i \implies t \bowtie c_{i+1}$ for all $i$. 
Thus, $t<c$ appears before any other predicate $t\bowtie c+\alpha$ for $\alpha \geq 0$. 
For simplicity, let us assume that $t$ is a single variable.
To enable a binary representation we find disjoint intervals representing the same constraints on variable values. 
Namely, replace the predicates in $P_t$ with 
\begin{inparaenum}[(1)]
\item
$t\bowtie c_0$,
\item
for $0<i\leq n$ the predicate $\neg (t\bowtie c_{i-1}) \wedge t\bowtie c_i$, and finally, 
\item
$\neg(t\bowtie c_n)$.
\end{inparaenum}
Effectively, forming a partition of the real line $\mathbb{R}$.

Let $part(P_t) = \{ t\bowtie c_0, \neg(t\bowtie c_{i-1})\wedge t\bowtie c_i, \neg(t\bowtie c_n) ~|~ 0<i\leq n\}$.
We call the left- and right-most partitions the \emph{border} partitions since they capture the left and right intervals to infinity. 
The other formulas define non-intersecting bounded intervals/partitions along $\mathbb{R}$. Fig.~\ref{fig:encoding} illustrates these partitions:
this set of formulas covers the whole line, i.e. for each point $t = c$, there is a formula $f$ in $part(P_t)$ such that $(t = c) \models f$. Further, note how each two distinct formulas $f_1, f_2 \in part(P_t)$ are mutually exclusive. Namely, $f_1\wedge f_2\equiv\bot$. 
Given this mutual exclusivity, it is easy to construct a  representation to reduce the number of binary variables in the predicate abstraction. 
The complexity of computing these partitions is only the complexity of sorting $P_t$ in ascending order based on values.

In a standard predicate abstraction approach, the number of predicates is ${{\sum}}_{t \in \textit{terms}} |P_{t}|$. 
With this encoding, they shrink to ${{\sum}}_{t \in \textit{terms}} \lceil|log_2(|P_t | + 1)|\rceil$.
\begin{figure}[t]
\centering
\resizebox{0.7\textwidth}{!}{%
\begin{tikzpicture}
\tikzstyle{every node}=[font=\LARGE]
\node [font=\LARGE] at (3.5,3.75) {$t \leq c_0$};
\node [font=\LARGE] at (5.1,5.25) {$\neg(t \leq c_0) \wedge t \leq c_1$};
\node [font=\LARGE] at (11.5,5.25) {$\neg(t \leq c_{n-1}) \wedge t \leq c_n$};
\node [font=\LARGE] at (13.2,3.75) {$\neg(t \leq c_n)$};
\draw [-{Stealth[length=3mm, width=3mm]}] (3.5,7.5) -- (14.5,7.5);
\draw [-{Stealth[length=3mm, width=3mm]}] (13,7) -- (2.25,7);
\node [font=\LARGE] at (5,6.5) {$GF t_{dec} \Rightarrow (GF t_{inc}) \vee t \leq c_0$};
\node [font=\LARGE] at (11.75,8) {$GF t_{inc} \Rightarrow (GF t_{dec}) \vee \neg(t \leq c_n)$};
\draw [-] (2.25,3.25) -- (14.5,3.25);
\draw [-] (4.75,3.25) -- (4.75,3);
\draw [-] (6.75,3.25) -- (6.75,3);
\draw [-] (10,3.25) -- (10,3);
\draw [-] (12,3.25) -- (12,3);
\node [font=\LARGE] at (4.75,2.5) {$c_0$};
\node [font=\LARGE] at (6.75,2.5) {$c_1$};
\draw [line width=0.9pt, -] (4.75,4.25) -- (2.25,4.25);
\draw[fill=black] (4.75,4.25) circle (0.15);
\draw[fill=black] (2.25,4.25) circle (0.15);
\draw [line width=0.9pt, -] (4.75,4.75) -- (6.75,4.75);
\draw[fill=black] (6.75,4.75) circle (0.15);
\draw[fill=white] (4.75,4.75) circle (0.15);
\draw [line width=0.9pt, -] (10,4.75) -- (12,4.75);
\draw[fill=white] (10,4.75) circle (0.15);
\draw[fill=black] (12,4.75) circle (0.15);
\draw [line width=0.9pt, -] (12,4.25) -- (14.5,4.25);
\draw[fill=white] (12,4.25) circle (0.15);
\draw[fill=black] (14.5,4.25) circle (0.15);

\node [font=\LARGE] at (2.25,2.5) {-$\infty$};
\node [font=\LARGE] at (14.5,2.5) {$\infty$};
\node [font=\LARGE] at (10,2.5) {$c_{n-1}$};
\node [font=\LARGE] at (12,2.5) {$c_n$};
\node [font=\LARGE] at (8.3,4.75) {...};
\draw [-] (2.25,3.25) -- (2.25,3);
\draw [-] (14.5,3.25) -- (14.5,3);
\end{tikzpicture}
}%
\vspace*{-2mm}
\caption{Partitions for binary encoding.}
\label{fig:encoding}
\end{figure}
Moreover, this enables a more efficient predicate abstraction computation: given we know each formula in $part(P_t)$ is mutually exclusive, we can consider each formula separately. Then, for each $t$ instead of performing $2^{2\times|P_t|}$ satisfiability checks we just need $(|P_t| + 1)^2$, giving a polynomial time complexity in terms of predicates, $({{\prod}}_{t \in terms} (|P_t | + 1))^2$, instead of the exponential $2^{2 \times {{\sum}}_{t \in \textit{terms}} |P_{t}|}$. 
The complexity of synthesis improves very significantly in terms of predicates, to $2^{{\prod}_{t \in \textit{terms}} |P_{t}| + 1}$, instead of $2^{2^{{\sum}_{t \in \textit{terms}} |P_{t}|}}$. 

Note that, to get the full view of time complexity for both abstraction and synthesis, the complexity described must be respectively multiplied by $|dom(\delta)| \times 2^{|B|}$ and $2^{2^{|B|}}$, where $B$ is the set of Boolean propositions in the concrete problem.

As an optimisation, if both terms $t$ and $-t$ are part of the abstraction, we transform predicates over $-t$ to predicates over $t$: $-t \leq c$ becomes $t \geq -c$, which becomes $\neg(t < -c)$. 
We note the approach described applies to both LIA and LRA, and might have applications beyond our approach.

\smallskip\noindent\textit{Acceleration.}
The partitioning optimises the encoding of predicates extracted from the problem and learned from safety refinements. 
Moreover, it allows to identify liveness properties relevant to the infinite-state arena. 

Consider that an abstract execution is within the leftmost partition, e.g., within $t \leq 0$. 
An increment in $t$ in the arena leads to an environment choice in the abstraction of whether to stay within $t\leq 0$ or move to the next partition.
Suppose the controller can repeatedly increment $t$ with a value bounded from $0$.

In the abstraction, the environment can still force an abstract execution satisfying $t \leq 0$ forever. 
The same is true for every partition, unless its size is smaller than the increment, e.g., a partition with one element. 
This abstract behaviour is not concretisable. 
That is, for every concrete value of $t$ and every $c$, after a finite number of increments bounded from $0$, the predicate $t\bowtie c$ becomes false. 
Similarly for any other partition. The dual is true for decrements. 
We note that in LIA, every increment or decrement is bounded from $0$. 

We encode this fact using fairness assumptions that rely on detecting increases and decreases of a term's value with transition predicates.
If for a term $t$ we identify that all changes of $t$ in $\arena$ are at least $\epsilon$, we define the transition predicates $t_{inc} := t_{prev} \leq t-\epsilon$ and $t_{dec} := t \leq t_{prev}-\epsilon$, refining the abstraction by a memory of when transitions increase or decrease the value of $t$. 
Notice that as changes to $t$ are at least $\epsilon$, when both $t_{dec}$ and $t_{inc}$ are false $t$ does not change. 
We then add the fairness assumptions: $(GF t_{dec}) \mathord{\implies} GF (t_{inc} \vee f_l)$ and $(GF t_{inc}) \mathord{\implies} GF (t_{dec} \vee f_r)$, where $f_l$ ($f_r$) is $t$'s left-(right-)most partitions.

The first (second) assumption enforces every abstract execution where $t$ strictly decreases (increases) and does not increase (decrease), to make progress towards the left-(right-)most partition. Thus, the environment cannot block the controller from exiting a partition, if they can repeatedly force a bounded from~0 decrease (increase) without increases (decreases).
For each term, we can then add these two corresponding fairness LTL assumptions to the abstraction. 
If the left- and right-most partitions are updated during safety refinement, we update the predicates inside these fairness assumptions with the new border partitions, ensuring we only ever have at most two such assumptions per term. In our implementation for LIA $\epsilon = 1$, and to optimise we leave out these assumptions if we cannot identify increases or decreases bounded from $0$ in the arena.

\section{Evaluation}\label{sec:eval}

\newcommand*{\TIMEOUT}{--}
\newcommand*{\TO}{\TIMEOUT}
\newcommand*{\DIVERGENCE}{$\inf$}
\newcommand*{\ERROR}{unk}
\newcommand*{\unk}{unk}
\newcommand*{\ERR}{\ERROR}
\newcommand*{\WRONGRESULT}{err}
\newcommand*{\NOTSUPPORTED}{n/a}

We implemented this approach in 
a tool\footnote{\url{https://github.com/shaunazzopardi/sweap}. An artifact for this paper is available~\cite{di_stefano_2025_15189175}.} targeting discrete synthesis problems. State-of-the-art tools are used as sub-routines: Strix~\cite{DBLP:conf/cav/MeyerSL18} (\ltl synthesis), nuXmv~\cite{DBLP:conf/cav/CavadaCDGMMMRT14} (invariant checking), MathSAT~\cite{DBLP:conf/tacas/CimattiGSS13} (interpolation and SMT checking), and CPAchecker~\cite{DBLP:conf/cav/BeyerK11} (termination checking). As a further optimisation, the tool performs also a binary encoding of the states variables of the arena, given they are mutually exclusive.

\begin{figure}[t]
\centering
\subfloat[Synthesis.\label{fig:performance synthesis}]{%
\includegraphics[width=0.45\textwidth]{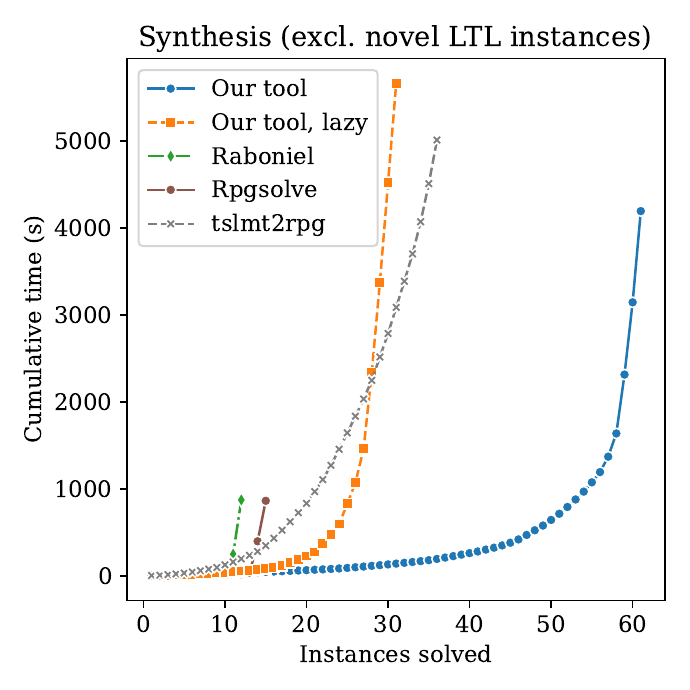}}
\subfloat[Realisability.\label{fig:performance realisability}]{\includegraphics[width=0.55\textwidth]{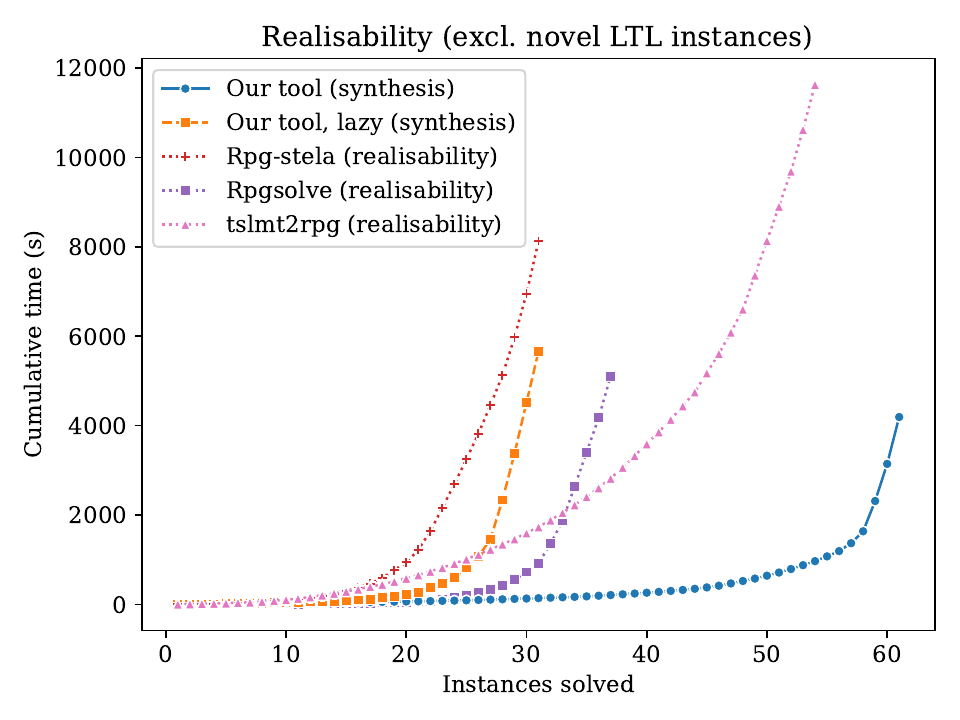}}%

\caption{Time comparison.}
\end{figure}

We compare our tool against 5 tools from literature
\texttt{raboniel}~\cite{MaderbacherBloem22}, \texttt{temos}~\cite{10.1145/3519939.3523429}, \texttt{rpgsolve}~\cite{10.1145/3632899}, \texttt{rpg-STeLA}~\cite{DBLP:conf/cav/SchmuckHDN24}, and \texttt{tslmt2rpg} (+\texttt{rpgsolve})~\cite{DBLP:journals/pacmpl/HeimD25}. We consider also a purely lazy version of our tool, with acceleration turned off to evaluate its utility. 
We do not compare against other tools fully outperformed by the \texttt{rpg} tools~\cite{DBLP:journals/corr/abs-2306-02427,DBLP:journals/corr/abs-2007-03656}, limited to safety/reachability~\cite{10.1007/978-3-030-81685-8_42,10.1145/3158149,8894254}, and another we could not acquire~\cite{DBLP:conf/isola/MaderbacherWB24}. 
All experiments ran
on a Linux workstation equipped with 32~GiB of memory and an Intel i7-5820K~CPU,
under a time limit of 20 minutes and a memory limit of 16 GiB. We show cumulative synthesis times in Fig.~\ref{fig:performance synthesis} for tools that support synthesis, and cumulative realisability times 
for other tools compared with our tools' cumulative synthesis times in Fig.~\ref{fig:performance realisability}.
\todo{S: graphs have different scale, doesn t look the best}

\begin{toappendix}
\subsection{Description of New Benchmarks}\label{apx:new-benchmarks}
\np{Without space restrictions should be extended in the polishing stage.}
\lds{added description of robot-tasks, please check.}
In \texttt{robot-tasks}, the environment gives the controller tasks of two kinds.
To do so, it initially sets two counters $x, y$, representing the number of tasks of each kind, to arbitrary natural numbers.
Then, the controller must perform all $x$-tasks (decrementing $x$ until it reaches~0) and then is allowed to perform a single $y$-task.
When this happens, the environment is again allowed to set $x$ to any positive value. The controller's goal is to eventually decrement $y$ to 0.
In \texttt{arbiter}, repeatedly, the environment makes a number of requests; then, the controller has to emit the same number of grants, which however may be delayed before succeeding. 
\texttt{arbiter-failure} is a variant where each grant may potentially fail: failures are controlled by the environment, under the constraint that grants must succeed infinitely often.
\texttt{elevator} is the example from Fig.~\ref{fig:example}. 
In \texttt{rep-reach-obst-xd}, the controller has to repeatedly reach a target
set by the environment, while the environment is allowed to set obstacles. These obstacles
hinder the progress of the controller, which must navigate around them.
The \texttt{rev-lane} problem describes a reversible traffic lane whose entry points can be shut or opened by the controller. Traffic initially flows in one direction.
Whenever the environment asks to change the flow, the controller must eventually effect the change without risking a car crash. The unrealisable version (\texttt{rev-lane-u}) allows cars to not exit the lane.
\texttt{robot-collect-v4} is a variation on the \texttt{robot-collect} examples from~\cite{DBLP:conf/cav/SchmuckHDN24}.
In the original benchmarks, a robot needs to get to a mine, collect a required number of samples there, and then bring the samples back. The robot needs to repeat this infinitely often. The robot was ensured that it would find at least one sample, even if the environment did not collaborate.
We remove this assumption and let the environment decide how many samples (if any) the robot collects at each time step;
however, we do impose a weaker fairness requirement on the environment, according to which it will supply samples infinitely often.
Lastly,
\texttt{taxi-service} (and its unrealisable variant, \texttt{taxi-service-u}) extend \texttt{elevator} to 2D space: the environment repeatedly sets a target location that the taxi must reach, and may also add obstacles to delay progress.
\end{toappendix}

\begin{toappendix}

\subsection{Further Details on Experimental Evaluation}\label{apx:other-tools}

\smallskip
\noindent\emph{System Configuration.}
Before running our experiments, we slightly tweaked our machine's operating system (Ubuntu 22.04 running Linux 5.15.0) in an attempt to make time measurements more uniform. Specifically, we set \emph{swappiness} to 0, forced a high-performance frequency governor for all CPU cores, and disabled simultaneous multithreading. Experiments were run sequentially.

\begin{table}[tb]
    \centering
    \tiny
    \caption{%
    Experimental evaluation on
    \textbf{RPG}solve,
    \textbf{R}PG-\textbf{St}eLA,
    \textbf{T}slmt\textbf{2R}pg,
    \textbf{Rab}oniel,
    \textbf{Tem}os,
    and our \textbf{S}ynthesis tool
    (with and without \emph{acc}eleration).
    \textbf{U}~marks unrealisable instances.
    \TIMEOUT\ denotes timeout, %
    \ERROR\ an error or inconclusive result,
    \textsf{x} an incorrect verdict.
    The best synthesis times are set in bold. 
    }\label{tab:results}
    
\begin{tabular}{|c|lr|c||c|c|c||c|c|c|c||c|c|}\hline
\multirow{2}{*}{G.}
& \multirow{2}{*}{Name, source} &
& \multirow{2}{*}{U}
& \multicolumn{3}{c||}{Realisability (s)}
& \multicolumn{6}{c|}{Synthesis (s)}\\\cline{5-13}
& & & & RPG & T2R & RSt & RPG & T2R & Rab & Tem & S$_{\textit{acc}}$ & S\\\hline\hline
\multirow{11}{*}{\rotatebox[origin=c]{90}{Safety}}
&  \textsf{box} & \cite{10.1145/3632899} &  & 0.24 & 53.82 & 23.48 & \textbf{0.57} & 92.20 & 1.30 & \TIMEOUT & 11.29 & 6.95 \\
&  \textsf{box-limited} & \cite{10.1145/3632899} &  & 0.23 & 10.58 & 5.19 & \textbf{0.38} & 15.36 & 0.41 & \textsf{x} & 3.35 & 2.88 \\
&  \textsf{diagonal} & \cite{10.1145/3632899} &  & 0.24 & 49.65 & 6.23 & \textbf{0.39} & 35.54 & 7.17 & \textsf{x} & 3.10 & 2.93 \\
&  \textsf{evasion} & \cite{10.1145/3632899} &  & 0.24 & 110.85 & 39.40 & \textbf{0.63} & 191.99 & 4.23 & \TIMEOUT & 16.51 & 8.03 \\
&  \textsf{follow} & \cite{10.1145/3632899} &  & 0.36 & \TIMEOUT & \TIMEOUT & \textbf{0.79} & \TIMEOUT & \TIMEOUT & \TIMEOUT & \ERROR & 875.44 \\
&  \textsf{solitary} & \cite{10.1145/3632899} &  & 0.18 & 7.96 & 1.39 & 0.37 & 9.05 & \textbf{0.34} & \textsf{x} & 3.56 & 3.48 \\
&  \textsf{square} & \cite{10.1145/3632899} &  & 0.24 & 243.63 & 174.40 & \textbf{0.59} & 370.69 & 90.30 & \TIMEOUT & 70.75 & 33.97 \\
&  \textsf{g-real} & \cite{DBLP:journals/pacmpl/HeimD25} &  & 10.36 & 305.21 & \TIMEOUT & \TIMEOUT & 301.28 & \textbf{4.20} & \TIMEOUT & 7.70 & 4.83 \\
&  \textsf{g-unreal-1} & \cite{DBLP:journals/pacmpl/HeimD25} & $\bullet$ & \TIMEOUT & 28.90 & \TIMEOUT & \TIMEOUT & \textbf{30.62} & \TIMEOUT & \ERROR & 266.94 & 244.15 \\
&  \textsf{g-unreal-2} & \cite{DBLP:journals/pacmpl/HeimD25} & $\bullet$ & 7.24 & 75.25 & \TIMEOUT & \textbf{6.98} & 40.84 & \ERROR & \ERROR & \ERROR & \ERROR \\
&  \textsf{g-unreal-3} & \cite{DBLP:journals/pacmpl/HeimD25} & $\bullet$ & \TIMEOUT & 44.51 & \TIMEOUT & \TIMEOUT & \textbf{44.54} & \TIMEOUT & \ERROR & \TIMEOUT & \TIMEOUT \\
\hline\hline
\multirow{25}{*}{\rotatebox[origin=c]{90}{Reachability}}
&  \textsf{heim-double-x} & \cite{10.1145/3632899} &  & 0.82 & 133.02 & 9.73 & 374.30 & 502.06 & \TIMEOUT & \textsf{x} & \textbf{78.41} & 102.11 \\
&  \textsf{robot-cat-real-1d} & \cite{10.1145/3632899} &  & 47.90 & \TIMEOUT & \TIMEOUT & \TIMEOUT & \TIMEOUT & \TIMEOUT & \TIMEOUT & \textbf{9.69} & 231.93 \\
&  \textsf{robot-cat-unreal-1d} & \cite{10.1145/3632899} & $\bullet$ & 42.88 & \TIMEOUT & 127.76 & \TIMEOUT & \TIMEOUT & \TIMEOUT & \ERROR & 9.14 & \textbf{8.38} \\
&  \textsf{robot-cat-real-2d} & \cite{10.1145/3632899} &  & \TIMEOUT & \TIMEOUT & \TIMEOUT & \TIMEOUT & \TIMEOUT & \TIMEOUT & \TIMEOUT & \textbf{677.28} & \TIMEOUT \\
&  \textsf{robot-cat-unreal-2d} & \cite{10.1145/3632899} & $\bullet$ & \TIMEOUT & \TIMEOUT & \TIMEOUT & \TIMEOUT & \TIMEOUT & \ERROR & \TIMEOUT & 830.81 & \textbf{95.98} \\
&  \textsf{robot-grid-reach-1d} & \cite{10.1145/3632899} &  & 0.31 & 5.96 & 1.23 & \textbf{1.02} & 10.05 & \TIMEOUT & \textsf{x} & 2.47 & 8.72 \\
&  \textsf{robot-grid-reach-2d} & \cite{10.1145/3632899} &  & 0.45 & 41.14 & 3.81 & \TIMEOUT & \ERROR & \ERROR & \textsf{x} & \textbf{3.62} & 44.87 \\
&  \textsf{sort4} & \cite{DBLP:conf/isola/MaderbacherWB24} &  & \ERROR & 477.29 & \ERROR & \ERROR & \ERROR & 624.27 & \textsf{x} & 107.09 & \textbf{42.80} \\
&  \textsf{sort5} & \cite{DBLP:conf/isola/MaderbacherWB24} &  & \ERROR & \TIMEOUT & \ERROR & \ERROR & \TIMEOUT & \TIMEOUT & \textsf{x} & \TIMEOUT & \TIMEOUT \\
&  \textsf{F-G-contradiction-1} & \cite{DBLP:journals/pacmpl/HeimD25} & $\bullet$ & \TIMEOUT & 32.58 & \TIMEOUT & \TIMEOUT & \textbf{32.69} & \TIMEOUT & \ERROR & \ERROR & \TIMEOUT \\
&  \textsf{F-G-contradiction-2} & \cite{DBLP:journals/pacmpl/HeimD25} & $\bullet$ & \TIMEOUT & 137.20 & \TIMEOUT & \TIMEOUT & 137.25 & \textbf{0.18} & \ERROR & 2.60 & 2.28 \\
&  \textsf{f-real} & \cite{DBLP:journals/pacmpl/HeimD25} &  & \TIMEOUT & 64.47 & \TIMEOUT & \TIMEOUT & 67.06 & \ERROR & \textsf{x} & \textbf{9.46} & 1039.40 \\
&  \textsf{f-unreal} & \cite{DBLP:journals/pacmpl/HeimD25} & $\bullet$ & \TIMEOUT & 104.75 & \TIMEOUT & \TIMEOUT & 106.98 & \TIMEOUT & \ERROR & 2.22 & \textbf{2.02} \\
&  \textsf{ordered-visits} & \cite{DBLP:journals/pacmpl/HeimD25} &  & \TIMEOUT & \TIMEOUT & \TIMEOUT & \TIMEOUT & \TIMEOUT & \TIMEOUT & \TIMEOUT & \textbf{3.72} & \ERROR \\
&  \textsf{ordered-visits-choice} & \cite{DBLP:journals/pacmpl/HeimD25} &  & \TIMEOUT & \TIMEOUT & \TIMEOUT & \TIMEOUT & \TIMEOUT & \TIMEOUT & \TIMEOUT & \textbf{2.76} & \ERROR \\
&  \textsf{precise-reachability} & \cite{DBLP:journals/pacmpl/HeimD25} &  & \TIMEOUT & \TIMEOUT & \TIMEOUT & \TIMEOUT & 12.33 & \TIMEOUT & \textsf{x} & \textbf{3.24} & 15.38 \\
&  \textsf{robot-to-target} & \cite{DBLP:journals/pacmpl/HeimD25} &  & \TIMEOUT & 423.61 & \TIMEOUT & \TIMEOUT & \textbf{437.56} & \TIMEOUT & \TIMEOUT & \ERROR & \TIMEOUT \\
&  \textsf{robot-to-target-unreal} & \cite{DBLP:journals/pacmpl/HeimD25} & $\bullet$ & \TIMEOUT & 312.40 & \TIMEOUT & \TIMEOUT & \textbf{313.70} & \TIMEOUT & \TIMEOUT & \TIMEOUT & \ERROR \\
&  \textsf{robot-to-target-charging} & \cite{DBLP:journals/pacmpl/HeimD25} &  & \TIMEOUT & 268.00 & \TIMEOUT & \TIMEOUT & \textbf{299.19} & \ERROR & \TIMEOUT & 1048.91 & \TIMEOUT \\
&  \textsf{robot-to-target-charging-unreal} & \cite{DBLP:journals/pacmpl/HeimD25} & $\bullet$ & \TIMEOUT & 18.88 & \TIMEOUT & \TIMEOUT & \textbf{17.41} & \TIMEOUT & \TIMEOUT & \TIMEOUT & \TIMEOUT \\
&  \textsf{thermostat-F} & \cite{DBLP:journals/pacmpl/HeimD25} &  & \TIMEOUT & 89.74 & \TIMEOUT & \TIMEOUT & 97.28 & \TIMEOUT & \ERROR & \textbf{4.53} & 1143.68 \\
&  \textsf{thermostat-F-unreal} & \cite{DBLP:journals/pacmpl/HeimD25} & $\bullet$ & \TIMEOUT & 164.55 & \TIMEOUT & \TIMEOUT & \textbf{165.19} & \TIMEOUT & \ERROR & \ERROR & \ERROR \\
&  \textsf{unordered-visits-charging} & \cite{DBLP:journals/pacmpl/HeimD25} &  & \TIMEOUT & \TIMEOUT & \TIMEOUT & \TIMEOUT & \TIMEOUT & \TIMEOUT & \TIMEOUT & \ERROR & \TIMEOUT \\
&  \textsf{unordered-visits} & \cite{DBLP:journals/pacmpl/HeimD25} &  & \TIMEOUT & 213.46 & \TIMEOUT & \TIMEOUT & 214.05 & \ERROR & \ERROR & \textbf{8.16} & 124.35 \\
&  \textsf{robot-tasks} &  &  & \TIMEOUT & \TIMEOUT & \TIMEOUT & \TIMEOUT & \TIMEOUT & \TIMEOUT & \textsf{x} & \textbf{3.07} & \TIMEOUT \\
\hline\hline
\multirow{45}{*}{\rotatebox[origin=c]{90}{Deterministic B\"uchi}}
&  \textsf{heim-buechi} & \cite{10.1145/3632899} &  & 2.65 & \TIMEOUT & \TIMEOUT & \TIMEOUT & \TIMEOUT & \ERROR & \textsf{x} & \textbf{3.58} & 1144.90 \\
&  \textsf{heim-fig7} & \cite{10.1145/3632899} & $\bullet$ & \TIMEOUT & 19.19 & \TIMEOUT & \TIMEOUT & 18.71 & \textbf{1.04} & \ERROR & 2.50 & 2.32 \\
&  \textsf{robot-commute-1d} & \cite{10.1145/3632899} &  & 1.22 & 1007.88 & 15.74 & \TIMEOUT & \TIMEOUT & \TIMEOUT & \textsf{x} & \textbf{4.27} & \TIMEOUT \\
&  \textsf{robot-commute-2d} & \cite{10.1145/3632899} &  & 7.52 & \ERROR & \ERROR & \TIMEOUT & \ERROR & \TIMEOUT & \TIMEOUT & \textbf{53.97} & \TIMEOUT \\
&  \textsf{robot-resource-1d} & \cite{10.1145/3632899} & $\bullet$ & 1.96 & \TIMEOUT & 6.07 & \textbf{2.01} & \TIMEOUT & \TIMEOUT & \ERROR & 7.81 & 10.19 \\
&  \textsf{robot-resource-2d} & \cite{10.1145/3632899} & $\bullet$ & 2.92 & \ERROR & 19.12 & \textbf{2.98} & \ERROR & \TIMEOUT & \ERROR & \TIMEOUT & \TIMEOUT \\
&  \textsf{chain-4} & \cite{DBLP:conf/cav/SchmuckHDN24} &  & 40.84 & 40.84 & 179.26 & \ERROR & \ERROR &  &  & \textbf{26.08} & \TIMEOUT \\
&  \textsf{chain-5} & \cite{DBLP:conf/cav/SchmuckHDN24} &  & 69.54 & 69.54 & 631.84 & \ERROR & \ERROR &  &  & \textbf{177.12} & \TIMEOUT \\
&  \textsf{chain-6} & \cite{DBLP:conf/cav/SchmuckHDN24} &  & 117.65 & 117.65 & \TIMEOUT & \ERROR & \ERROR &  &  & \TIMEOUT & \TIMEOUT \\
&  \textsf{chain-7} & \cite{DBLP:conf/cav/SchmuckHDN24} &  & 177.70 & 177.70 & \TIMEOUT & \ERROR & \ERROR &  &  & \ERROR & \TIMEOUT \\
&  \textsf{chain-simple-5} & \cite{DBLP:conf/cav/SchmuckHDN24} &  & 17.65 & 17.65 & 28.81 & \TIMEOUT & \TIMEOUT &  &  & \textbf{3.57} & \TIMEOUT \\
&  \textsf{chain-simple-10} & \cite{DBLP:conf/cav/SchmuckHDN24} &  & 54.40 & 54.40 & 85.94 & \TIMEOUT & \TIMEOUT &  &  & \textbf{5.07} & \TIMEOUT \\
&  \textsf{chain-simple-20} & \cite{DBLP:conf/cav/SchmuckHDN24} &  & 196.66 & 196.66 & 276.12 & \TIMEOUT & \TIMEOUT &  &  & \textbf{8.87} & \TIMEOUT \\
&  \textsf{chain-simple-30} & \cite{DBLP:conf/cav/SchmuckHDN24} &  & 434.02 & 434.02 & 575.16 & \TIMEOUT & \TIMEOUT &  &  & \textbf{14.34} & \ERROR \\
&  \textsf{chain-simple-40} & \cite{DBLP:conf/cav/SchmuckHDN24} &  & 764.40 & 764.40 & 974.14 & \TIMEOUT & \TIMEOUT &  &  & \textbf{21.83} & \ERROR \\
&  \textsf{chain-simple-50} & \cite{DBLP:conf/cav/SchmuckHDN24} &  & \TIMEOUT & \TIMEOUT & \TIMEOUT & \TIMEOUT & \TIMEOUT &  &  & \textbf{30.93} & \TIMEOUT \\
&  \textsf{chain-simple-60} & \cite{DBLP:conf/cav/SchmuckHDN24} &  & \TIMEOUT & \TIMEOUT & \TIMEOUT & \TIMEOUT & \TIMEOUT &  &  & \textbf{37.99} & \TIMEOUT \\
&  \textsf{chain-simple-70} & \cite{DBLP:conf/cav/SchmuckHDN24} &  & \TIMEOUT & \TIMEOUT & \TIMEOUT & \TIMEOUT & \TIMEOUT &  &  & \textbf{51.51} & \TIMEOUT \\
&  \textsf{items-processing} & \cite{DBLP:conf/cav/SchmuckHDN24} &  & 101.39 & 101.39 & 535.70 & \TIMEOUT & \TIMEOUT &  &  & \textbf{17.09} & \TIMEOUT \\
&  \textsf{robot-analyze} & \cite{DBLP:conf/cav/SchmuckHDN24} &  & 931.22 & 931.22 & 79.51 & \TIMEOUT & \TIMEOUT &  &  & \textbf{8.71} & \TIMEOUT \\
&  \textsf{robot-collect-v1} & \cite{DBLP:conf/cav/SchmuckHDN24} &  & 769.76 & 769.76 & 17.35 & \TIMEOUT & \TIMEOUT &  &  & \textbf{4.34} & \TIMEOUT \\
&  \textsf{robot-collect-v2} & \cite{DBLP:conf/cav/SchmuckHDN24} &  & \ERROR & \ERROR & 418.33 & \TIMEOUT & \TIMEOUT &  &  & \textbf{4.38} & \ERROR \\
&  \textsf{robot-collect-v3} & \cite{DBLP:conf/cav/SchmuckHDN24} &  & 769.24 & 769.24 & 45.62 & \TIMEOUT & \TIMEOUT &  &  & \textbf{17.70} & \TIMEOUT \\
&  \textsf{robot-deliver-v1} & \cite{DBLP:conf/cav/SchmuckHDN24} &  & \TIMEOUT & \TIMEOUT & 73.68 & \TIMEOUT & \TIMEOUT &  &  & \textbf{8.17} & \TIMEOUT \\
&  \textsf{robot-deliver-v2} & \cite{DBLP:conf/cav/SchmuckHDN24} &  & \TIMEOUT & \TIMEOUT & 553.22 & \TIMEOUT & \TIMEOUT &  &  & \textbf{21.81} & \TIMEOUT \\
&  \textsf{robot-deliver-v3} & \cite{DBLP:conf/cav/SchmuckHDN24} &  & \TIMEOUT & \TIMEOUT & 849.01 & \TIMEOUT & \TIMEOUT &  &  & \textbf{86.25} & \TIMEOUT \\
&  \textsf{robot-deliver-v4} & \cite{DBLP:conf/cav/SchmuckHDN24} &  & \TIMEOUT & \TIMEOUT & \TIMEOUT & \TIMEOUT & \TIMEOUT &  &  & \textbf{54.08} & \TIMEOUT \\
&  \textsf{robot-deliver-v5} & \cite{DBLP:conf/cav/SchmuckHDN24} &  & \TIMEOUT & \TIMEOUT & \TIMEOUT & \TIMEOUT & \TIMEOUT &  &  & \textbf{64.48} & \TIMEOUT \\
&  \textsf{robot-repair} & \cite{DBLP:conf/cav/SchmuckHDN24} &  & \TIMEOUT & \TIMEOUT & \TIMEOUT & \TIMEOUT & \TIMEOUT &  &  & \TIMEOUT & \TIMEOUT \\
&  \textsf{robot-running} & \cite{DBLP:conf/cav/SchmuckHDN24} &  & 515.37 & 515.37 & 520.04 & \TIMEOUT & \TIMEOUT &  &  & \textbf{18.78} & \TIMEOUT \\
&  \textsf{scheduler} & \cite{DBLP:conf/cav/SchmuckHDN24} &  & 6.54 & 6.54 & 1181.35 & \TIMEOUT & \TIMEOUT &  &  & \textbf{3.60} & 35.37 \\
&  \textsf{buffer-storage} & \cite{DBLP:journals/pacmpl/HeimD25} &  & 5.67 & \TIMEOUT & \TIMEOUT & 465.27 & \TIMEOUT & 138.04 & \ERROR & \textbf{5.47} & 6.16 \\
&  \textsf{gf-real} & \cite{DBLP:journals/pacmpl/HeimD25} &  & \TIMEOUT & 2.92 & \TIMEOUT & \TIMEOUT & 3.28 & \TIMEOUT & \textsf{x} & \textbf{2.09} & \TIMEOUT \\
&  \textsf{gf-unreal} & \cite{DBLP:journals/pacmpl/HeimD25} & $\bullet$ & 2.20 & 5.99 & \TIMEOUT & 2.12 & 2.50 & \textbf{0.21} & \ERROR & \TIMEOUT & \TIMEOUT \\
&  \textsf{GF-G-contradiction} & \cite{DBLP:journals/pacmpl/HeimD25} & $\bullet$ & \TIMEOUT & 6.38 & \TIMEOUT & \TIMEOUT & \textbf{6.44} & \TIMEOUT & \ERROR & \TIMEOUT & \TIMEOUT \\
&  \textsf{helipad} & \cite{DBLP:journals/pacmpl/HeimD25} &  & \ERROR & 152.74 & \ERROR & \ERROR & 184.20 & \ERROR & \TIMEOUT & \textbf{6.27} & 18.85 \\
&  \textsf{helipad-contradict} & \cite{DBLP:journals/pacmpl/HeimD25} & $\bullet$ & 3.05 & 264.16 & 672.97 & \textbf{3.12} & 269.79 & \TIMEOUT & \ERROR & 117.76 & 5.41 \\
&  \textsf{package-delivery} & \cite{DBLP:journals/pacmpl/HeimD25} &  & \ERROR & 116.61 & \TIMEOUT & \ERROR & 135.16 & \ERROR & \TIMEOUT & \textbf{90.00} & \ERROR \\
&  \textsf{patrolling} & \cite{DBLP:journals/pacmpl/HeimD25} &  & \TIMEOUT & 265.88 & \TIMEOUT & \TIMEOUT & \textbf{268.96} & \TIMEOUT & \TIMEOUT & \ERROR & \ERROR \\
&  \textsf{patrolling-alarm} & \cite{DBLP:journals/pacmpl/HeimD25} &  & \TIMEOUT & 83.49 & \TIMEOUT & \TIMEOUT & \textbf{84.94} & \TIMEOUT & \ERROR & \TIMEOUT & \textsf{x} \\
&  \textsf{storage-GF-64} & \cite{DBLP:journals/pacmpl/HeimD25} &  & \TIMEOUT & \TIMEOUT & \TIMEOUT & \TIMEOUT & \TIMEOUT & \TIMEOUT & \TIMEOUT & \textbf{2.94} & 6.31 \\
&  \textsf{tasks} & \cite{DBLP:journals/pacmpl/HeimD25} &  & \TIMEOUT & 786.23 & \TIMEOUT & \TIMEOUT & \TIMEOUT & \TIMEOUT & \ERROR & \textbf{3.51} & 388.61 \\
&  \textsf{tasks-unreal} & \cite{DBLP:journals/pacmpl/HeimD25} & $\bullet$ & \TIMEOUT & 181.58 & \TIMEOUT & \TIMEOUT & \textbf{188.37} & \TIMEOUT & \ERROR & \ERROR & \ERROR \\
&  \textsf{thermostat-GF} & \cite{DBLP:journals/pacmpl/HeimD25} &  & \TIMEOUT & 275.86 & \TIMEOUT & \TIMEOUT & 198.39 & \TIMEOUT & \ERROR & \textbf{15.52} & \ERROR \\
&  \textsf{thermostat-GF-unreal} & \cite{DBLP:journals/pacmpl/HeimD25} & $\bullet$ & \TIMEOUT & 80.90 & \TIMEOUT & \TIMEOUT & \textbf{104.35} & \TIMEOUT & \ERROR & \TIMEOUT & \TIMEOUT \\
\hline
\end{tabular}

\end{table}

\end{toappendix}

\newcommand*{\COMPEXPERIMENTS}{81}
\newcommand*{\LITERATUREEXPERIMENTS}{80}
\newcommand*{\LTLEXPERIMENTS}{15}
\newcommand*{\SWEAPSCORE}{61}
\newcommand*{\SWEAPLAZYSCORE}{31}
\newcommand*{\SECONDBEST}{tslmt2rpg}
\newcommand*{\SECONDBESTSCORE}{36}
\newcommand*{\LAZYBEST}{3}
\newcommand*{\NOREFINEMENTS}{11}

\smallskip
\noindent
\textit{Benchmarks.}
We collect \LITERATUREEXPERIMENTS{} LIA benchmarks from the literature. Most encode practical problems, such as robotic mission control, job scheduling, sorting, or data buffering. They are defined in TSL~\cite{10.1007/978-3-030-25540-4_35} or as deterministic games,
and may include arbitrary integers as input, which we equivalently encode with extra steps that
let the environment set variables to any finite value
(see Section~\ref{sec:disc}).
All these benchmarks consist of problems encodable as deterministic B\"uchi games.
Some benchmarks~\cite{DBLP:conf/cav/SchmuckHDN24} 
compose multiple such games together, for added difficulty. 
{Following others, we ignore problems~\cite{10.1145/3519939.3523429,10.1007/978-3-030-25540-4_35} that are trivial.}
We only introduce one novel reachability 
game to these benchmarks, \texttt{robot-tasks},\footnote{Appendix~\ref{apx:new-benchmarks} has more details about this new benchmark.}
that we crafted to highlight the limitations of previous approaches compared to our own.
Some of the problems from~\cite{DBLP:conf/cav/SchmuckHDN24} are not available in TSL format. We test those on neither \texttt{raboniel} nor \texttt{temos} but we expect they would both
fail, as their techniques are insufficient
for B\"uchi goals (see Section~\ref{sec:disc}), and for \texttt{tslmt2rpg} %
we simply consider the time taken by \texttt{rpgsolve} on the corresponding RPG problem.

\smallskip
\noindent
\textit{Results (comparative evaluation).}\footnote{Appendix~\ref{apx:more-data} has additional experimental data, and an extended discussion.}
It is clear from Fig.~\ref{fig:performance synthesis} that the eager version of our tool
solves almost double more synthesis problems than the best competitor, and faster. 
The lazy version is comparable to the best competitor.
For realisability, Fig.~\ref{fig:performance realisability}
shows our tool with acceleration scaling and performing
much better on synthesis than the other tools do on realisability. However, the lazy version is outperformed by the rpg tools. %
Table~\ref{tab:summ-tables} summarises the evaluation;
for each tool we report the number of solved problems (out of \COMPEXPERIMENTS{}), the ones it solved in the shortest time, and those no other tool was able to solve.
Our tool is the clear winner in each category.
If we consider synthesis, even without acceleration we are comparable to the state of the art:
our tools solve \SWEAPSCORE{} (eager) and \SWEAPLAZYSCORE{} (lazy) problems, while the best competitor
\begin{table}[t]
\caption{Experimental results.}
\subfloat[Comparative evaluation of
\textbf{Rab}oniel, \textbf{Tem}os, \textbf{RPG}solve, \textbf{T}slmt\textbf{2R}pg, \textbf{R}pg-\textbf{St}eLa, and our \textbf{S}yn\-the\-sis tool, with and without \textit{acc}eleration.\label{tab:summ-tables}]{\parbox{0.53\textwidth}{\footnotesize
\begin{tabular}{|p{5em}||c|c|c|c||c|c|}\hline
Synthesis & Rab & Tem & RPG & T2R & S$_{\textit{acc}}$ & S \\\hline
    solved & 12 & 0 & 15 & 36 & \textbf{61} & 31\\
    best & 5 & 0 & 11 & 13 & \textbf{43} & 4\\
    unique & 0 & 0 & 1 & 11 & \textbf{27} & 0\\\hline
\end{tabular}\\
\begin{tabular}{|p{6.2em}||c|c|c||c|c|}\hline
Realisability & RPG & RSt & T2R & S$_{\textit{acc}}$ & S \\\hline
    solved & 37 & 31 & 54 & \textbf{61} & 31\\
    best & 21 & 0 & 13 & \textbf{37} & 7\\
    unique & 0 & 0 & \textbf{11} & 9 & 0\\\hline
\end{tabular}
}}
\hspace{1em}
\subfloat[LTL benchmarks.\label{tab:ltl}]{
\parbox{0.47\textwidth}{\footnotesize\begin{tabular}{|c|c||c|c|}
\hline
\multirow{2}{*}{Name} & \multirow{2}{*}{U} & \multicolumn{2}{c|}{Time (s)}\\\cline{3-4}
& & S$_{\textit{acc}}$ & S\\\hline\hline
\textsf{arbiter} & {} & \textbf{2.77} & 4.90\\\hline
\textsf{arbiter-failure} & {} & 2.04 & \textbf{1.98}\\\hline
\textsf{elevator} & {} & \textbf{2.53} & 15.92\\\hline
\textsf{infinite-race} & {} & \textbf{1.98} & 4.38\\\hline
\textsf{infinite-race-u} & {$\bullet$} & \TIMEOUT & \TIMEOUT\\\hline
\textsf{infinite-race-unequal-1} & {} & \textbf{6.50} & \TIMEOUT\\\hline
\textsf{infinite-race-unequal-2} & {} & \TIMEOUT & \TIMEOUT\\\hline
\textsf{reversible-lane-r} & {} & \textbf{7.39} & 17.53\\\hline
\textsf{reversible-lane-u} & {$\bullet$} & 18.70 & \textbf{4.54}\\\hline
\textsf{rep-reach-obst-1d} & {} & \textbf{2.47} & 9.04\\\hline
\textsf{rep-reach-obst-2d} & {} & \textbf{3.85} & 38.51\\\hline
\textsf{rep-reach-obst-6d} & {} & \TIMEOUT & \TIMEOUT\\\hline
\textsf{robot-collect-v4} & {} & \textbf{16.51} & \TIMEOUT\\\hline
\textsf{taxi-service} & {} & \textbf{39.26} & 68.02\\\hline
\textsf{taxi-service-u} & {$\bullet$} & 4.14 & \textbf{3.50}\\\hline

\end{tabular}
}}
\end{table}
\texttt{\SECONDBEST{}} solves \SECONDBESTSCORE{}.
When looking closely at the behaviour on the easiest
instances (see Fig.~\ref{fig:cactus-easy}), we see that our tool has an initialization overhead of a few seconds 
while other tools can solve simple problems in under 1s. However, our tool scales better. 
We also ran our lazy tool without the binary encoding, and measured noticeably worse performances:
it times out on two more problems, and takes on average 10\% more time (see Fig.~\ref{fig:speedup}).%
\todo{more discussion about this; how does acceleration without binarisation perform?}

\smallskip\noindent
\textit{Evaluation on novel LTL benchmarks.}
{%
We contribute \LTLEXPERIMENTS{} benchmarks with \ltl objectives unencodable as deterministic B\"uchi objectives, i.e., they are theoretically out of scope for other tools. For sanity checking we attempted them on the other tools and validated their inability to decide these problems. We do not include them with the previous benchmarks to ensure a fairer evaluation. Three of these benchmarks could be solved by other tools if infinite-range inputs are used (\texttt{arbiter}, \texttt{infinite-race}, and \texttt{infinite-race-u}), but they fail since incrementing and decrementing requires environment fairness constraints.

These benchmarks involve control of cyber-physical systems
such as the elevator from Fig.~\ref{fig:example}, variations thereof, a reversible traffic lane, and robotic missions, some of which are extensions of literature benchmarks.
They also include strong fairness and/or let the environment delay progress for the controller.\footnote{These benchmarks are also described in detail in Appendix~\ref{apx:new-benchmarks}.}
}
Table~\ref{tab:ltl} reports how both configurations of our
tool handle our novel benchmarks.
Column U marks unrealisable problems.
The lazy approach outperforms the eager one on just \LAZYBEST{} benchmarks out of \LTLEXPERIMENTS{}.
On \NOREFINEMENTS{} problems, acceleration enriches 
the first abstraction enough to lead immediately to a verdict. We note that solving \texttt{infinite-race-unequal-1} requires structural refinement, as it allows infinite amount of increments and decrements, but of unequal value, while for literature benchmarks acceleration is enough.

\smallskip
\noindent
\textit{Failure Analysis.}
Lastly, we discuss four limitations in our approach exposed by our experiments. Section~\ref{sec:disc} contains more detail on when and why the other tools fail.
The first is inherent to synthesis: the Boolean synthesis problem may become big enough to exceed machine resources. A bespoke finite-state synthesis procedure could mitigate this, by relying on the underlying parity game
rather than creating fresh problems.

The second is that some unrealisable problems admit no finite \counterstrategies in our setting. \texttt{robot-repair}, which no tool solves, is the only such example from literature (we also designed \texttt{infinite-race-u} to be of this kind). Briefly, this involves two stages: a losing loop for which the controller controls exit and (after the loop) a state wherein the goal is unreachable. The environment cannot universally 
quantify over all predicates (since it controls them), hence no finite 
\counterstrategy exists. But if we construct the dual problem,  by swapping objectives between the environment and controller,
we do find a strategy for the original environment goal.
We are working on automating this dualisation.
    
The third is that our requirements for when to apply structural refinement may be too strong, and thus some loops go undiscovered. Instead of looking for loops solely in the counterexample prefix, one may instead consider the strongly connected components of the \counterstrategy.

Lastly, there are pathological counterexamples, irrelevant to the problem, 
that involve the controller causing an incompatibility by going to a partition and the environment not being able to determine exactly when dec/increments should force an exit from this partition. This is the main cause of failure for our lazy approach. Modifications to concretisability checking might avoid this issue.

\begin{toappendix}
\smallskip
\noindent\emph{Additional experimental data.}\label{apx:more-data}
Table~\ref{tab:results} (see penulminate page) contains non-aggregated data from our comparative evaluation.
Figure~\ref{fig:cactus-easy} contains a more detailed view (in log-scale) of what is the bottom-left
corner of Figures~\ref{fig:performance synthesis} and~\ref{fig:performance realisability}.
{We notice that \texttt{rpgsolve} is clearly the fastest to determine realisability on the
simplest benchmarks (the safety problems from~\cite{8894254}), but its performance on synthesis
degrades sharply. \texttt{tslmt2rpg} and \texttt{raboniel} show
similar performance issues, and none of them manages to reach past 20 solved benchmarks.}
Our tools are somewhat slower at first but do scale more gracefully, with the eager
configuration of our tool showing the slowest trend in runtime increase.

Table~\ref{tab:details} (see last page) contains details about the number of state predicates,
transition predicates, and the number of refinements 
performed by each configuration of our tool \sweap for each benchmark.
Column \textit{acc} indicates whether the row refers to the acceleration
configuration \sweap$\!\!_\textit{acc}$ or not.
Then, column
\textit{init} reports the number of initial
state predicates (\textit{s}) and transition predicates (\textit{p}).
Note, $x$ initial transition predicates indicates $x/2$
accelerations performed, i.e. $x/2$ strong fairness constraints
added as assumptions to the abstract LTL problem.
The \textit{ref} column indicates the number of safety refinement (\textit{sf}) and structural loop refinements (\textit{sl}) performed,
and finally \textit{add} indicates the number of state predicates (\textit{sp})
and transition predicates (\textit{tp}) added by such refinements.
Fig.~\ref{fig:speedup} is a scatter plot that compares the execution times on benchmarks successfully solved by
our lazy tool, with and without the binary encoding from Section~\ref{sec:efficient}. Each dot is a benchmark;
dots falling above (below) the diagonal represent problems where the tool with binary encoding is faster (slower)
than the baseline. Most dots fall above the line, and indeed we measure an average 1.10x speedup with a maximum
of 2.02x. Even in the worst case, the solver with binary encoding is only 0.94x as fast as the one without,
which may be explained by the slight overhead to compute the encoding possibly combined with normal fluctuations
in execution times.

\begin{figure}[t]
    \centering
    \includegraphics[width=0.8\linewidth]{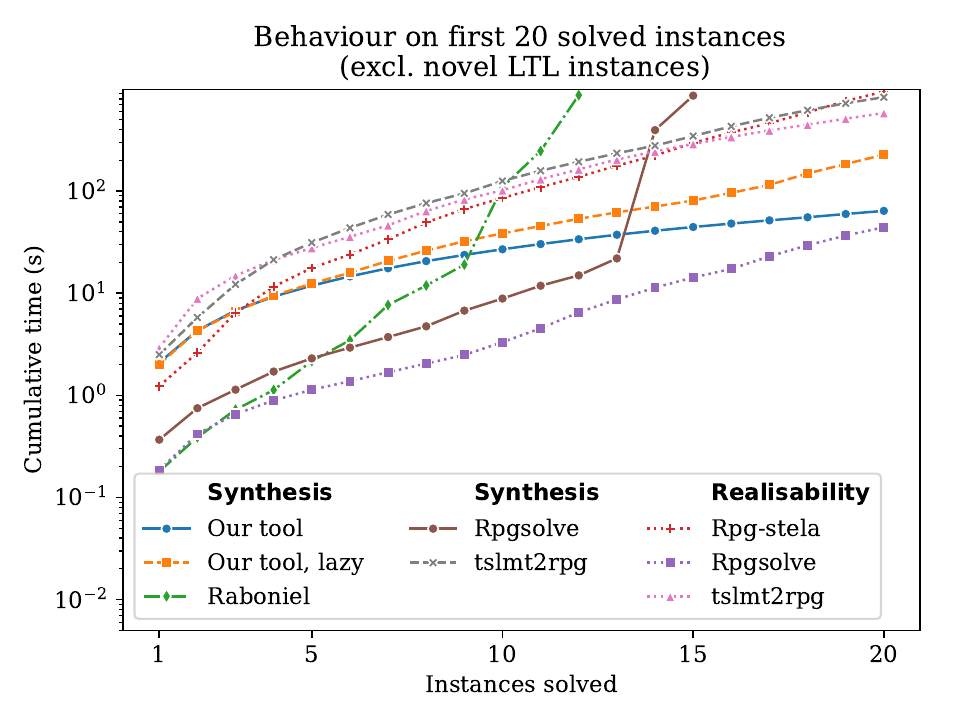}
    \caption{Experimental evaluation on the quickest 20 problems.}\label{fig:cactus-easy}
\end{figure}

\begin{figure}[t]
    \centering
    \includegraphics[width=0.5\linewidth]{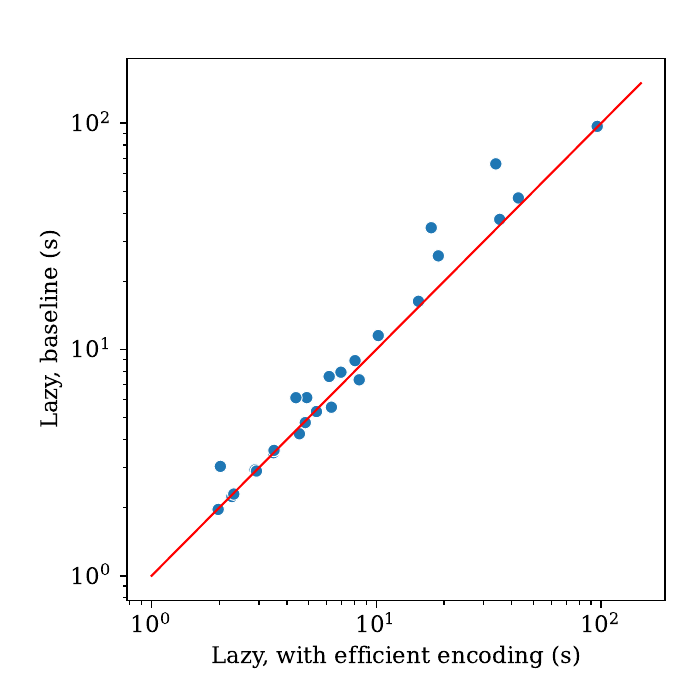}
    \caption{Scatter plot showing the speedup provided by the binary encoding from Section~\ref{sec:efficient}.}\label{fig:speedup}
\end{figure}

\newcounter{n}
\begin{table}[h]
\caption{Experiment details for \sweap{$_{acc}$} and \sweap.}\label{tab:details}
\centering\tiny
\begin{tabular}[t]{|l||c||c|c|c|c|c|c||}
\hline
&& \multicolumn{2}{c|}{init}
& \multicolumn{2}{c|}{ref}
& \multicolumn{2}{c||}{add}\\\hline
\multicolumn{1}{|c||}{Name} & acc & s & t &sf. &sl. & sp & tp\\\hline\hline\multirow{2}{*}[0em]{arbiter}

& $\bullet$
& 3 & 2 & 0 & 0 & 0 & 0\\\cline{2-8}
& 
& 3 & 0 & 0 & 1 & 0 & 2\\\hline\multirow{2}{*}[0em]{arbiter-failure}

& $\bullet$
& 2 & 2 & 0 & 0 & 0 & 0\\\cline{2-8}
& 
& 2 & 0 & 0 & 0 & 0 & 0\\\hline\multirow{2}{*}[0em]{box}

& $\bullet$
& 2 & 2 & 2 & 0 & 2 & 0\\\cline{2-8}
& 
& 2 & 0 & 1 & 0 & 2 & 0\\\hline\multirow{2}{*}[0em]{box-limited}

& $\bullet$
& 2 & 2 & 0 & 0 & 0 & 0\\\cline{2-8}
& 
& 2 & 0 & 0 & 0 & 0 & 0\\\hline\multirow{2}{*}[0em]{buffer-storage}

& $\bullet$
& 6 & 4 & 0 & 0 & 0 & 0\\\cline{2-8}
& 
& 6 & 0 & 1 & 0 & 2 & 0\\\hline\multirow{2}{*}[0em]{chain-4}

& $\bullet$
& 10 & 8 & 0 & 0 & 0 & 0\\\cline{2-8}
& 
& 10 & 0 & 6 & 0 & 6 & 0\\\hline\multirow{2}{*}[0em]{chain-5}

& $\bullet$
& 12 & 10 & 0 & 0 & 0 & 0\\\cline{2-8}
& 
& 12 & 0 & 3 & 0 & 3 & 0\\\hline\multirow{2}{*}[0em]{chain-6}

& $\bullet$
& 14 & 12 & 0 & 0 & 0 & 0\\\cline{2-8}
& 
& 14 & 0 & 1 & 0 & 1 & 0\\\hline\multirow{2}{*}[0em]{chain-7}

& $\bullet$
& 16 & 14 & 0 & 0 & 0 & 14\\\cline{2-8}
& 
& 16 & 0 & 0 & 0 & 0 & 0\\\hline\multirow{2}{*}[0em]{chain-simple-10}

& $\bullet$
& 4 & 2 & 0 & 0 & 0 & 0\\\cline{2-8}
& 
& 4 & 0 & 5 & 1 & 6 & 4\\\hline\multirow{2}{*}[0em]{chain-simple-20}

& $\bullet$
& 4 & 2 & 0 & 0 & 0 & 0\\\cline{2-8}
& 
& 4 & 0 & 4 & 1 & 4 & 4\\\hline\multirow{2}{*}[0em]{chain-simple-30}

& $\bullet$
& 4 & 2 & 0 & 0 & 0 & 0\\\cline{2-8}
& 
& 4 & 0 & 4 & 1 & 4 & 4\\\hline\multirow{2}{*}[0em]{chain-simple-40}

& $\bullet$
& 4 & 2 & 0 & 0 & 0 & 0\\\cline{2-8}
& 
& 4 & 0 & 4 & 1 & 4 & 4\\\hline\multirow{2}{*}[0em]{chain-simple-5}

& $\bullet$
& 4 & 2 & 0 & 0 & 0 & 0\\\cline{2-8}
& 
& 4 & 0 & 5 & 1 & 6 & 4\\\hline\multirow{2}{*}[0em]{chain-simple-50}

& $\bullet$
& 4 & 2 & 0 & 0 & 0 & 0\\\cline{2-8}
& 
& 4 & 0 & 4 & 1 & 4 & 4\\\hline\multirow{2}{*}[0em]{chain-simple-60}

& $\bullet$
& 4 & 2 & 0 & 0 & 0 & 0\\\cline{2-8}
& 
& 4 & 0 & 4 & 0 & 4 & 0\\\hline\multirow{2}{*}[0em]{chain-simple-70}

& $\bullet$
& 4 & 2 & 0 & 0 & 0 & 0\\\cline{2-8}
& 
& 4 & 0 & 4 & 0 & 4 & 0\\\hline\multirow{2}{*}[0em]{diagonal}

& $\bullet$
& 2 & 2 & 0 & 0 & 0 & 0\\\cline{2-8}
& 
& 2 & 0 & 0 & 0 & 0 & 0\\\hline\multirow{2}{*}[0em]{elevator}

& $\bullet$
& 2 & 2 & 0 & 0 & 0 & 0\\\cline{2-8}
& 
& 2 & 0 & 0 & 2 & 0 & 4\\\hline\multirow{2}{*}[0em]{evasion}

& $\bullet$
& 4 & 4 & 1 & 0 & 2 & 0\\\cline{2-8}
& 
& 4 & 0 & 1 & 0 & 1 & 0\\\hline\multirow{2}{*}[0em]{F-G-contradiction-1}

& $\bullet$
& 0 & 0 & 0 & 0 & 0 & 0\\\cline{2-8}
& 
& 5 & 0 & 5 & 6 & 10 & 3\\\hline\multirow{2}{*}[0em]{F-G-contradiction-2}

& $\bullet$
& 2 & 4 & 0 & 0 & 0 & 0\\\cline{2-8}
& 
& 2 & 0 & 0 & 0 & 0 & 0\\\hline\multirow{2}{*}[0em]{f-real}

& $\bullet$
& 2 & 4 & 0 & 0 & 0 & 0\\\cline{2-8}
& 
& 2 & 0 & 1 & 1 & 2 & 8\\\hline\multirow{2}{*}[0em]{f-unreal}

& $\bullet$
& 1 & 2 & 0 & 0 & 0 & 0\\\cline{2-8}
& 
& 1 & 0 & 0 & 0 & 0 & 0\\\hline\multirow{2}{*}[0em]{follow}

& $\bullet$
& 8 & 8 & 2 & 0 & 3 & 8\\\cline{2-8}
& 
& 8 & 0 & 5 & 0 & 6 & 0\\\hline\multirow{2}{*}[0em]{g-real}

& $\bullet$
& 7 & 2 & 1 & 0 & 1 & 2\\\cline{2-8}
& 
& 7 & 0 & 1 & 0 & 1 & 0\\\hline\multirow{2}{*}[0em]{g-unreal-1}

& $\bullet$
& 1 & 2 & 21 & 0 & 42 & 0\\\cline{2-8}
& 
& 1 & 0 & 21 & 0 & 42 & 0\\\hline\multirow{2}{*}[0em]{g-unreal-2}

& $\bullet$
& 5 & 4 & 3 & 0 & 4 & 8\\\cline{2-8}
& 
& 5 & 0 & 5 & 0 & 9 & 0\\\hline\multirow{2}{*}[0em]{g-unreal-3}

& $\bullet$
& 3 & 2 & 13 & 3 & 23 & 3\\\cline{2-8}
& 
& 3 & 0 & 6 & 8 & 8 & 3\\\hline\multirow{2}{*}[0em]{GF-G-contradiction}

& $\bullet$
& 3 & 2 & 3 & 4 & 6 & 3\\\cline{2-8}
& 
& 3 & 0 & 3 & 6 & 4 & 3\\\hline\multirow{2}{*}[0em]{gf-real}

& $\bullet$
& 1 & 2 & 0 & 0 & 0 & 0\\\cline{2-8}
& 
& 1 & 0 & 1 & 1 & 2 & 2\\\hline\multirow{2}{*}[0em]{gf-unreal}

& $\bullet$
& 1 & 0 & 39 & 0 & 78 & 1\\\cline{2-8}
& 
& 1 & 0 & 40 & 0 & 80 & 0\\\hline\multirow{2}{*}[0em]{heim-buechi}

& $\bullet$
& 3 & 4 & 0 & 0 & 0 & 0\\\cline{2-8}
& 
& 3 & 0 & 21 & 1 & 42 & 2\\\hline\multirow{2}{*}[0em]{heim-double-x}

& $\bullet$
& 3 & 4 & 15 & 0 & 21 & 0\\\cline{2-8}
& 
& 3 & 0 & 12 & 1 & 21 & 2\\\hline\multirow{2}{*}[0em]{heim-fig7}

& $\bullet$
& 1 & 2 & 0 & 0 & 0 & 0\\\cline{2-8}
& 
& 1 & 0 & 0 & 0 & 0 & 0\\\hline\multirow{2}{*}[0em]{helipad}

& $\bullet$
& 2 & 2 & 0 & 0 & 0 & 0\\\cline{2-8}
& 
& 2 & 0 & 0 & 1 & 0 & 2\\\hline\multirow{2}{*}[0em]{helipad-contradict}

& $\bullet$
& 2 & 2 & 5 & 0 & 5 & 2\\\cline{2-8}
& 
& 2 & 0 & 0 & 0 & 0 & 0\\\hline\multirow{2}{*}[0em]{infinite-race}

& $\bullet$
& 1 & 2 & 0 & 0 & 0 & 0\\\cline{2-8}
& 
& 1 & 0 & 0 & 1 & 0 & 3\\\hline\multirow{2}{*}[0em]{infinite-race-u}

& $\bullet$
& 1 & 2 & 10 & 3 & 11 & 4\\\cline{2-8}
& 
& 1 & 0 & 21 & 5 & 21 & 4\\\hline\multirow{2}{*}[0em]{infinite-race-unequal-1}

& $\bullet$
& 1 & 2 & 0 & 1 & 0 & 4\\\cline{2-8}
& 
& 1 & 0 & 57 & 0 & 59 & 0\\\hline\multirow{2}{*}[0em]{infinite-race-unequal-2}

& $\bullet$
& 1 & 2 & 2 & 6 & 4 & 5\\\cline{2-8}
& 
& 1 & 0 & 16 & 3 & 17 & 5\\\hline\multirow{2}{*}[0em]{items-processing}

& $\bullet$
& 7 & 4 & 0 & 0 & 0 & 0\\\cline{2-8}
& 
& 7 & 0 & 4 & 1 & 8 & 3\\\hline\multirow{2}{*}[0em]{ordered-visits}

& $\bullet$
& 2 & 2 & 2 & 0 & 2 & 0\\\cline{2-8}
& 
& 2 & 0 & 1 & 0 & 0 & 0\\\hline\multirow{2}{*}[0em]{ordered-visits-choice}

& $\bullet$
& 3 & 2 & 1 & 0 & 1 & 0\\\cline{2-8}
& 
& 3 & 0 & 1 & 0 & 0 & 0\\\hline\multirow{2}{*}[0em]{package-delivery}

& $\bullet$
& 4 & 2 & 0 & 0 & 0 & 0\\\cline{2-8}
& 
& 4 & 0 & 1 & 1 & 1 & 10\\\hline\multirow{2}{*}[0em]{patrolling}

& $\bullet$
& 8 & 2 & 15 & 0 & 30 & 0\\\cline{2-8}
& 
& 8 & 0 & 3 & 2 & 4 & 2\\\hline\multirow{2}{*}[0em]{patrolling-alarm}

& $\bullet$
& 8 & 2 & 10 & 0 & 20 & 0\\\cline{2-8}
& 
& 8 & 0 & 1 & 1 & 0 & 2\\\hline\multirow{2}{*}[0em]{precise-reachability}

& $\bullet$
& 1 & 2 & 1 & 0 & 1 & 1\\\cline{2-8}
& 
& 1 & 0 & 0 & 1 & 0 & 4\\\hline
\end{tabular}\begin{tabular}[t]{|l||c||c|c|c|c|c|c||}
\hline
&& \multicolumn{2}{c|}{init}
& \multicolumn{2}{c|}{ref}
& \multicolumn{2}{c||}{add}\\\hline
\multicolumn{1}{|c||}{Name} & acc & s & t &sf. &sl. & sp & tp\\\hline\hline\multirow{2}{*}[0em]{rep-reach-obst-1d}

& $\bullet$
& 2 & 2 & 0 & 0 & 0 & 0\\\cline{2-8}
& 
& 2 & 0 & 0 & 2 & 0 & 3\\\hline\multirow{2}{*}[0em]{rep-reach-obst-2d}

& $\bullet$
& 4 & 4 & 0 & 0 & 0 & 0\\\cline{2-8}
& 
& 4 & 0 & 0 & 4 & 0 & 6\\\hline\multirow{2}{*}[0em]{rep-reach-obst-6d}

& $\bullet$
& 12 & 12 & 0 & 0 & 0 & 0\\\cline{2-8}
& 
& 12 & 0 & 0 & 0 & 0 & 0\\\hline\multirow{2}{*}[0em]{reversible-lane-r}

& $\bullet$
& 4 & 4 & 0 & 0 & 0 & 0\\\cline{2-8}
& 
& 4 & 0 & 0 & 2 & 0 & 4\\\hline\multirow{2}{*}[0em]{reversible-lane-u}

& $\bullet$
& 4 & 4 & 2 & 0 & 2 & 0\\\cline{2-8}
& 
& 4 & 0 & 0 & 0 & 0 & 0\\\hline\multirow{2}{*}[0em]{robot-cat-real-1d}

& $\bullet$
& 4 & 4 & 0 & 0 & 0 & 0\\\cline{2-8}
& 
& 4 & 0 & 2 & 2 & 2 & 3\\\hline\multirow{2}{*}[0em]{robot-cat-real-2d}

& $\bullet$
& 8 & 8 & 0 & 0 & 0 & 0\\\cline{2-8}
& 
& 8 & 0 & 3 & 0 & 3 & 0\\\hline\multirow{2}{*}[0em]{robot-cat-unreal-1d}

& $\bullet$
& 4 & 4 & 0 & 0 & 0 & 0\\\cline{2-8}
& 
& 4 & 0 & 0 & 0 & 0 & 0\\\hline\multirow{2}{*}[0em]{robot-cat-unreal-2d}

& $\bullet$
& 8 & 8 & 0 & 0 & 0 & 0\\\cline{2-8}
& 
& 8 & 0 & 0 & 0 & 0 & 0\\\hline\multirow{2}{*}[0em]{robot-commute-1d}

& $\bullet$
& 4 & 4 & 0 & 0 & 0 & 0\\\cline{2-8}
& 
& 4 & 0 & 3 & 5 & 3 & 4\\\hline\multirow{2}{*}[0em]{robot-commute-2d}

& $\bullet$
& 8 & 8 & 0 & 0 & 0 & 0\\\cline{2-8}
& 
& 8 & 0 & 4 & 1 & 4 & 3\\\hline\multirow{2}{*}[0em]{robot-grid-reach-1d}

& $\bullet$
& 2 & 2 & 0 & 0 & 0 & 0\\\cline{2-8}
& 
& 2 & 0 & 0 & 2 & 0 & 3\\\hline\multirow{2}{*}[0em]{robot-grid-reach-2d}

& $\bullet$
& 4 & 4 & 0 & 0 & 0 & 0\\\cline{2-8}
& 
& 4 & 0 & 0 & 4 & 0 & 6\\\hline\multirow{2}{*}[0em]{robot-resource-1d}

& $\bullet$
& 2 & 3 & 2 & 0 & 4 & 0\\\cline{2-8}
& 
& 2 & 0 & 2 & 1 & 4 & 2\\\hline\multirow{2}{*}[0em]{robot-resource-2d}

& $\bullet$
& 4 & 5 & 15 & 1 & 16 & 2\\\cline{2-8}
& 
& 4 & 0 & 2 & 4 & 3 & 5\\\hline\multirow{2}{*}[0em]{robot-tasks}

& $\bullet$
& 4 & 4 & 0 & 0 & 0 & 0\\\cline{2-8}
& 
& 4 & 0 & 56 & 1 & 56 & 2\\\hline\multirow{2}{*}[0em]{robot-to-target}

& $\bullet$
& 6 & 4 & 0 & 0 & 0 & 4\\\cline{2-8}
& 
& 6 & 0 & 0 & 1 & 2 & 6\\\hline\multirow{2}{*}[0em]{robot-to-target-charging}

& $\bullet$
& 7 & 8 & 26 & 0 & 52 & 0\\\cline{2-8}
& 
& 7 & 0 & 28 & 1 & 57 & 3\\\hline\multirow{2}{*}[0em]{robot-to-target-charging-unreal}

& $\bullet$
& 8 & 10 & 8 & 0 & 15 & 0\\\cline{2-8}
& 
& 8 & 0 & 8 & 0 & 16 & 0\\\hline\multirow{2}{*}[0em]{robot-to-target-unreal}

& $\bullet$
& 6 & 4 & 3 & 0 & 3 & 0\\\cline{2-8}
& 
& 6 & 0 & 0 & 0 & 0 & 0\\\hline\multirow{2}{*}[0em]{robot-analyze}

& $\bullet$
& 9 & 8 & 0 & 0 & 0 & 0\\\cline{2-8}
& 
& 9 & 0 & 2 & 5 & 4 & 3\\\hline\multirow{2}{*}[0em]{robot-collect-v1}

& $\bullet$
& 6 & 6 & 0 & 0 & 0 & 0\\\cline{2-8}
& 
& 6 & 0 & 3 & 3 & 6 & 3\\\hline\multirow{2}{*}[0em]{robot-collect-v2}

& $\bullet$
& 8 & 6 & 0 & 0 & 0 & 0\\\cline{2-8}
& 
& 8 & 0 & 3 & 3 & 5 & 3\\\hline\multirow{2}{*}[0em]{robot-collect-v3}

& $\bullet$
& 8 & 10 & 0 & 0 & 0 & 0\\\cline{2-8}
& 
& 8 & 0 & 3 & 3 & 5 & 6\\\hline\multirow{2}{*}[0em]{robot-collect-v4}

& $\bullet$
& 8 & 10 & 0 & 0 & 0 & 0\\\cline{2-8}
& 
& 8 & 0 & 2 & 3 & 5 & 3\\\hline\multirow{2}{*}[0em]{robot-deliver-v1}

& $\bullet$
& 9 & 8 & 0 & 0 & 0 & 0\\\cline{2-8}
& 
& 9 & 0 & 3 & 6 & 3 & 3\\\hline\multirow{2}{*}[0em]{robot-deliver-v2}

& $\bullet$
& 11 & 8 & 0 & 0 & 0 & 0\\\cline{2-8}
& 
& 11 & 0 & 2 & 3 & 2 & 3\\\hline\multirow{2}{*}[0em]{robot-deliver-v3}

& $\bullet$
& 13 & 8 & 0 & 0 & 0 & 0\\\cline{2-8}
& 
& 13 & 0 & 1 & 1 & 2 & 3\\\hline\multirow{2}{*}[0em]{robot-deliver-v4}

& $\bullet$
& 15 & 8 & 0 & 0 & 0 & 0\\\cline{2-8}
& 
& 15 & 0 & 1 & 1 & 1 & 3\\\hline\multirow{2}{*}[0em]{robot-deliver-v5}

& $\bullet$
& 15 & 8 & 0 & 0 & 0 & 0\\\cline{2-8}
& 
& 15 & 0 & 1 & 1 & 1 & 3\\\hline\multirow{2}{*}[0em]{robot-repair}

& $\bullet$
& 8 & 4 & 4 & 2 & 8 & 3\\\cline{2-8}
& 
& 8 & 0 & 3 & 6 & 6 & 3\\\hline\multirow{2}{*}[0em]{robot-running}

& $\bullet$
& 9 & 10 & 0 & 0 & 0 & 0\\\cline{2-8}
& 
& 9 & 0 & 3 & 2 & 5 & 3\\\hline\multirow{2}{*}[0em]{scheduler}

& $\bullet$
& 2 & 4 & 0 & 0 & 0 & 0\\\cline{2-8}
& 
& 2 & 0 & 1 & 2 & 1 & 5\\\hline\multirow{2}{*}[0em]{solitary}

& $\bullet$
& 2 & 2 & 1 & 0 & 2 & 0\\\cline{2-8}
& 
& 2 & 0 & 1 & 0 & 2 & 0\\\hline\multirow{2}{*}[0em]{sort4}

& $\bullet$
& 3 & 6 & 2 & 0 & 2 & 4\\\cline{2-8}
& 
& 3 & 0 & 2 & 0 & 2 & 0\\\hline\multirow{2}{*}[0em]{sort5}

& $\bullet$
& 4 & 8 & 3 & 0 & 3 & 6\\\cline{2-8}
& 
& 4 & 0 & 5 & 0 & 5 & 0\\\hline\multirow{2}{*}[0em]{square}

& $\bullet$
& 4 & 4 & 5 & 0 & 8 & 0\\\cline{2-8}
& 
& 4 & 0 & 4 & 0 & 8 & 0\\\hline\multirow{2}{*}[0em]{storage-GF-64}

& $\bullet$
& 8 & 2 & 0 & 0 & 0 & 0\\\cline{2-8}
& 
& 8 & 0 & 0 & 1 & 0 & 2\\\hline\multirow{2}{*}[0em]{tasks}

& $\bullet$
& 10 & 3 & 0 & 0 & 0 & 0\\\cline{2-8}
& 
& 10 & 0 & 12 & 1 & 22 & 2\\\hline\multirow{2}{*}[0em]{tasks-unreal}

& $\bullet$
& 8 & 4 & 1 & 0 & 2 & 0\\\cline{2-8}
& 
& 8 & 0 & 3 & 2 & 4 & 3\\\hline\multirow{2}{*}[0em]{taxi-service}

& $\bullet$
& 4 & 4 & 0 & 0 & 0 & 0\\\cline{2-8}
& 
& 4 & 0 & 0 & 4 & 0 & 8\\\hline\multirow{2}{*}[0em]{taxi-service-u}

& $\bullet$
& 4 & 4 & 0 & 0 & 0 & 0\\\cline{2-8}
& 
& 4 & 0 & 0 & 0 & 0 & 0\\\hline\multirow{2}{*}[0em]{thermostat-F}

& $\bullet$
& 10 & 4 & 0 & 0 & 0 & 0\\\cline{2-8}
& 
& 10 & 0 & 13 & 2 & 25 & 3\\\hline\multirow{2}{*}[0em]{thermostat-F-unreal}

& $\bullet$
& 0 & 0 & 0 & 0 & 0 & 0\\\cline{2-8}
& 
& 0 & 0 & 0 & 0 & 0 & 0\\\hline\multirow{2}{*}[0em]{thermostat-GF}

& $\bullet$
& 10 & 4 & 0 & 0 & 0 & 0\\\cline{2-8}
& 
& 10 & 0 & 2 & 2 & 4 & 10\\\hline\multirow{2}{*}[0em]{thermostat-GF-unreal}

& $\bullet$
& 10 & 4 & 2 & 0 & 3 & 2\\\cline{2-8}
& 
& 10 & 0 & 4 & 2 & 7 & 4\\\hline\multirow{2}{*}[0em]{unordered-visits}

& $\bullet$
& 6 & 4 & 0 & 0 & 0 & 0\\\cline{2-8}
& 
& 6 & 0 & 0 & 1 & 0 & 2\\\hline\multirow{2}{*}[0em]{unordered-visits-charging}

& $\bullet$
& 4 & 4 & 2 & 0 & 4 & 0\\\cline{2-8}
& 
& 4 & 0 & 4 & 5 & 6 & 3\\\hline
\end{tabular}

\end{table}

\smallskip\noindent
\textit{Extended discussion of other tools.}
Realisability-wise the \texttt{rpg} tools are quite capable.
Here, \texttt{rpgsolve} performs much better than in \texttt{rpg-STeLA}'s evaluation~\cite{DBLP:conf/cav/SchmuckHDN24}.
That evaluation, however, seems
to use a configuration of \texttt{rpg-STeLA} that mimics \texttt{rpgsolve}'s approach, whereas we use the latest version of \texttt{rpgsolve} from \texttt{tslmt2rpg}'s software artifact~\cite{heim_2024_13939202}.
We ran \texttt{rpg-STeLA} in its ``normal'' configuration, which was the best performer in the experimental evaluation from~\cite{DBLP:conf/cav/SchmuckHDN24}.
Surprisingly, \texttt{rpgsolve} is sometimes faster to synthesise than \texttt{rpg-STeLA} is to determine realisability.
We also note that the artifact for \texttt{tslmt2rpg}~\cite{heim_2024_13939202} mentions that these tools are highly affected by performance-based heuristics, which could also explain some differences.
\texttt{temos} synthesises fairness constraints divorced from the objectives of these more sophisticated games, and thus does not solve any of our problems.
\texttt{raboniel} typically diverges due to its safety-only refinement loop; \texttt{rpg} tools may fail during quantifier elimination or by divergence (e.g., \texttt{robot-tasks}).
\end{toappendix}

\section{Related Work}\label{sec:disc}

Before discussing related synthesis approaches, we note that Balaban, Pnueli, and Zuck describe a similar CEGAR approach for infinite-state model checking \cite{10.1007/11562436_1}. From counterexamples they discover ranking functions for terminating loops, and encode their well-foundedness in the underlying fair discrete system, similar to how we encode well-foundedness during acceleration. Our structural refinement is instead more localised to specific loops. We may benefit from the more general ranking abstraction, but it is often easier to prove termination of loops through loop variants rather than ranking functions, which do not admit the same encoding.
Interestingly, their approach is relatively complete, i.e. given the right ranking functions and state predicates the LTL property can be verified. We cannot say the same about our approach, given, as mentioned in the previous section, there are some unrealisable problems we cannot terminate on.

We discuss the exact differences between our setting and that of 
TSL synthesis~\cite{10.1007/978-3-030-25540-4_35} and RPG~\cite{10.1145/3632899}.
We then discuss infinite-state synthesis more generally. 

\smallskip
\noindent
\textit{TSL and RPG compared to our approach.}
We start by noting that, in the context of linear integer arithmetic, for every possible synthesis problem in TSL or RPG, we can effectively construct an equi-realisable problem in our setting 
(see Appendix~\ref{app:tsl and rpg to us} for the full details). 
In both TSL and RPG, variables are partitioned between inputs and outputs.
At each step of the game,
the environment sets values for all inputs (so, choosing among potentially infinitely-many or continuously-many candidate values in one step) and the controller responds by choosing among a finite set of deterministic updates to its own variables.
The environment also initialises \emph{all} variables.
Dually, in our setting, players only own Boolean variables and have only a finite set of choices. 
Then, infinite-range variables are updated based on the joint choice.
For all three, repeating single interactions ad-infinitum leads to traces that are either checked to satisfy an LTL formula (TSL and our setting) or to satisfy safety, reachability, or repeated reachability
w.r.t. certain locations in the arena/program (RPG). The restriction to finite-range updates hinders the applicability of our approach to linear real arithmetic, given the necessity of repeated uncountable choices there.
However, we expect the more novel parts of our approach (liveness refinements and acceleration) to still be applicable in this richer theory. 
Indeed, we define acceleration in a way that it is also applicable for LRA in Section~\ref{sec:efficient}.

\begin{toappendix}
\subsection{Realisability modulo arenas vs.~TSL and RPG}
\label{app:tsl and rpg to us}
We show how to create equirealisable problems starting from TSL and RPG.
We include the required definitions adjusted to our notations. 

\subsubsection{TSL}
Consider a finite set of variables $V$ partitioned to inputs $V_E$  and outputs $V_C$ (note that here these are general variables and not Boolean variables as in the inputs and outputs in our arenas). 
We identify a finite set of predicates 
$P\subseteq \preds(V)$ and a set of finite sets of updates $\{U_v\}_{v \in V_C}$ such that for every $v\in V_C$ we have $U_v\subseteq \mathcal{T}(V)$. 
Given a predicate $p$ and a valuation $\val\in \Val(V)$ we write $\val\models p$ for the case that the valuation $\val$ satisfies the predicate $p$. 
For $\val\in \Val(V)$, we write $\val{\Downarrow}_{V_E}$ for the valuation $\val'\in \Val(V_E)$ such that for every $v\in V_E$ we have $\val(v)=\val'(t)$. 
We use the similar notation $\val{\Downarrow}_{V_C}$.
For an update $u\in U_t$ and two valuation $\val_1,\val_2\in \Val(V)$, we are interested in $\val_2(v)=u(\val_1{\Downarrow}_{V_C}\cup \val_2{\Downarrow}_{V_E})$.
That is, if the update is applied to the values of $V_C$ from $\val_1$ and the values of $V_e$ from $\val_2$. 
Let $U=\bigcup_{v\in V_C} U_v$ and let 
$\Pi_U=\prod_{v\in V_C}U_v$, we treat every element in $\Pi_U$ as the combination  of the individual updates in it.
We write $(\val_1,\val_2)\models \vec{u}$, when for every $v\in V_C$ we have $\val_2(t_i)=u(\val_1{\Downarrow}_{V_C}\cup \val_2{\Downarrow}_{V_E})$, where $\vec{u}=\langle u_{v_1},\ldots, u_{v_k}\rangle$. 
A TSL formula over $P$ and $U$ is a formula in $LTL(P\cup U)$.
A trace $(\val_0,\vec{u}_0),(\val_1,\vec{u}_1),\ldots \in (\Val(V) \times \Pi_U)^\Omega$ is \emph{consistent} if for  every $i\geq 1$ we have $(\val_{i-1},\val_i)\models \vec{u}_{i}$. 
It satisfies the formula $\phi$ if there is a sequence $p_0,\ldots,\in (2^P)^\omega$ such that for every $i$ we have $\val_i\models \cbigwedge p_i$ and the trace $(p_0,\vec{u}_0),(p_1,\vec{u}_1),\ldots$ satisfies $\phi$. 
Notice that $\vec{u}_0$ does not play a role in the consistency of a trace. To make this clear we sometimes simply write $-$ instead.
Given a letter $\sigma\in\Sigma$ and a language $L\subseteq\Sigma^\omega$ let $\sigma \cdot L=\{\sigma \cdot w ~|~ w\in L]\}$.

\begin{definition}[TSL Synthesis]
A TSL formula $\phi$ over $P$ and $U$ is said to be realisable if and only if for every $\val\in \Val(V)$ there is a Mealy machine $C_\val$, with input $\Sigma_{in}=\Val(V_E)$ and output $\Sigma_{out} = \Val(V_C)\times \Pi_U$ such that every trace in $(\val,-)\cdot L(C_v)$ is consistent and satisfies $\phi$. The problem of synthesis is to construct the machines $\{C_\val\}_{\val\in \Val({V})}$.
\end{definition}

Consider a TSL formula $\phi$ over $P$ and $U$, where all variables range over integers. 
We define an arena that allows the environment to choose the initial valuation $\val$ and in every step choose the valuation of the variables in $V_E$. 
This can be done by incrementing or decrementing the variables one by one to their desired values. 
Let $\arena_\phi$ denote the arena $\langle V_\phi,v_0,\delta_\phi\rangle$, where $V_\phi=V\cup \{ s\} \cup \mathbb{E} \cup \mathbb{C}$, $\mathbb{E}=\{inc,dec,var_e,var_c\}$, where $var_e$ and $var_c$ are finite-range variables ranging over $|V_e|$ and $|V_c|$ ($inc$ and $dec$ are Boolean), $\mathbb{C}=U$, where every $u\in U$ is a Boolean variable, $s$ ranges over $\{e,c\}$, $v_0$ is an arbitrary valuation setting $s$ to $c$, and $\delta_\phi$ includes the following guard-update pairs.
We skip the details corresponding to the Boolean encoding of the finite range variables $var_e$ and $var_c$ and use freely $v_{var_s}$ to denote the variable identified by the variable $var_s$, where $s$ itself identifies either $var_c$ or $var_e$.
$$
\delta_\phi ~= \left \{ \left .
\begin{array}{r@{\quad \mapsto\quad }l}
(inc \wedge \neg dec) & v_{var_s}\inc\\
(\neg inc \wedge dec) & v_{var_s}\dec\\
(s=c \wedge (inc \iff dec)) & s=e\\
(s=e \wedge (inc \iff dec) \wedge \vec{u}) & \vec{u}
\end{array}
\right | \vec{u} \in \Pi_U \right \}
$$

It follows that the TSL formula $\phi$ needs to be evaluated only over the locations where $s=e$ at the point that the environmnet decides to set $inc\iff dec$. Let $eval$ denote the predicate $(s=e) \wedge (inc \iff  dec)$ and let $t(\phi)$ denote the following recursive transformation on TSL/LTL.
Notice that we use the notation $\mathbb{AP}$ to refer to the predicates and updates appearing in the TSL formula $\phi$. Predicates are handled directly in the LTL formula that is the target for synthesis modulo the arena.
That is, they are evaluated over the variables in the arena.
Dually, the symbols in $U$ appear as Boolean values in both the arena and the formula and their semantics does not play a role in the evaluation of the formula~--~they are treated as mere syntax (their semantics obviously plays a major role in the construction of the arena above).
$$
\begin{array}{l l}
t(p) := eval \wedge p & p\in \mathbb{AP}\\
t(\phi_1 \varpropto \phi_2) := t(\phi_1) \varpropto t(\phi_2) & \varpropto \in \{\vee, \wedge \} \\
t(\neg \phi_1) := \neg t(\phi_1) \\
t(X\phi_1) := (\neg eval) U (eval \wedge t(\phi_1)) \\
t(\phi_1 U \phi_2) := (eval \implies t(\phi_1)) U (eval \wedge t(\phi_2)
\end{array}
$$

\begin{lemma}
    For every TSL formula $\phi$ over $P$ and $U$, $\phi$ is realisable iff $((GF eval) \implies t(\phi))$ is realisable modulo $\arena_\phi$. 
    \label{lem:tsl equirealisable}
\end{lemma}

We note that the general form of $\arena_\phi$, as presented here, is tailored for explainability rather than efficiency. 
In practice, TSL formulas appearing in benchmarks allow for very efficient representations as synthesis modulo arenas.
Benchmarks usually include full initialisation of all/most variables and they use input variables quite rigidly. 
This allows to remove the need to include $var_c$ and valuation $s=c$. 
It also simplifies the updates to $var_e$ that happen when $s=e$
\begin{proof}
\begin{itemize}
    \item[$\Rightarrow$]
    Consider a TSL formula $\phi$ and assume that it is realized by the family of Mealy machines ${M_{\val}}_{\val \in\Val(V)}$, where $\val\in\Val({V})$ is the initial valuation and $M_{\val}=\langle S_\val, s_0^\val, \Sigma_{in},\Sigma_{out},\delta_\val\rangle$.
    We construct the Mealy machine corresponding to the union of all machines ${M_{\val}}_{\val \in \Val(V)}$ with additional states corresponding to the sequences of actions $inc\wedge \neg dec$ and $\neg inc \wedge dec$ applied by the environment. 
    Formally, the states of the Mealy machine $C'$ are $\Val(V) \cup \bigcup_{\val\in \Val(V)} (S_\val \times \Val(V_E))$, the input is $2^{\{inc,dec\}}\times Dom(var_e) \times Dom(var_c)$, and the output is $2^{U}$. We set $\vec{0}$ as the initial state of $C'$ and include the following transitions.
    Let $u_0$ be some arbitrary output.
    $$
    \begin{array}{l l l}
    \rightarrow = &
    \left \{ \begin{array}{c}(\val,i,u_0,\val')\\((s^w,\val),i,u_0,(s^w,\val'))\end{array} \left |
    \begin{array}{l}
    s^{w}\in S_{w}, \val\in \Val(V),i(inc)=\top,i(dec)=\bot,\\
    \val'=\val[v_{i(var_{\val(s)})}\rightarrow \val(v_{i(var_{\val(s)})})+1]
    \end{array} \right . \right \} & \cup \\
    & \left \{ \begin{array}{c}(\val,i,u_0,\val')\\((s^w,\val),i,u_0,(s^w,\val')) \end{array}\left |
    \begin{array}{l}
    s^w\in S_w, \val\in \Val(V), i(inc)=\bot, i(dec)=\top, \\
    \val'=\val[v_{i(var_{\val(s)})}\rightarrow \val(v_{i(var_{\val(s)})})-1]
    \end{array} \right . \right \} & \cup \\
    & \left \{ (\val,i,u_0,\val') \left |
    \begin{array}{l}
    \val\in \Val(V), \val(s)=c, i(inc)=i(dec),\\
    \val'=\val[s\rightarrow e]
    \end{array} \right . \right \} & \cup \\
    & \left \{ \begin{array}{c}(\val,i,u_0,(s_0^\val,\val))\\((s^w,\val),i,u,(s_1^w,\val')) \end{array} \left |
    \begin{array}{l}
    s^w\in S_w, \val\in \Val(V), \val(s)=e, i(inc)=i(dec), \\
    (s^w,\val{\Downarrow}_{V_e},u,s^w_1)\in \delta_w,\\
    \val{\Downarrow}_{V_E}=\val'{\Downarrow}_{V_E}, (\val,\val')\models u
    \end{array} \right . \right \} 
    \end{array}
    $$

    Consider a computation of $C'$.
    It starts in a state corresponding to a valuation of $V$. It then uses the inputs $var_c$, $inc$, and $dec$ to set the values of all the variables in $V_C$.
    If at some point, $inc$ and $dec$ agree on their values, it updates the variable $s$ to $e$.
    Then, it uses the inputs $var_e$, $inc$, and $dec$ to set the values of all the variables in $V_E$.
    If, again, $inc$ and $dec$ agree on their values, it updates its state to the initial state of $M_w$ for the valuation $w$ determined by the environment keeping the values of the variables.
    Then, it starts simulating $M_w$.
    Indeed, it keeps the state of $M_w$ while allowing the environment to update the values of $V_E$ by using 
    the inputs $var_e$, $inc$, and $dec$ to set the values of all the variables in $V_E$.
    If the environment makes $inc$ and $dec$ agree on their values, it takes a transition corresponding to the way $M_w$ handles the valuation on $V_E$ that is decided.
    According to this transition it sets the update, which keeps the values of $V_E$ unchanged (indeed they were just updated by the environment) and updates the variables in $V_c$ according to the chosen update.
    It follows that every computation of $C'$ corresponds to a computation that either ends in an infinite value-search phase or has infinitely many value-search phases.
    A computation that ends in an infinite value-search phase, 
    does not satisfy $GF\,\textit{eval}$ and hence satisfies the specification.
    Otherwise, the computation corresponds to an infinite computation of $M_w$ with stuttering steps corresponding to environment choosing the values in $V_E$ where $\neg eval$ holds.
    Hence, the computation satisfies $t(\phi)$. 
    \item[$\Leftarrow$]
    Let $C'$ be a Mealy machine solving the realisability modulo $A_\phi$ of $t(\phi)$.
    We construct by induction the machines $M_w$.
    Consider an initial valuation $w_0$ chosen by the environment.
    We can find the state $s_{w_0}$ of $C'$ reached after the environment plays $inc$, $dec$, $var_e$ and $var_c$ so as to set all variables to the values in $w_0$. 
    Consider a state $s$ of $M_{w_0}$, which corresponds to a state $s$ of $C'$. 
    Then, for every environment choice $\val \in \Val(V_E)$, we can find the state $s_\val$ of $C'$ reached from $s$ after the environment plays $inc$, $dec$, $var_e$ to set all input variables to the values in $\val$.
    Then, if the environment plays $inc\iff dec$, then $C'$ takes a transition $(s_\val,i,u_\val,s')$. 
    We add to $M_{w_0}$ the transition $(s,\val{\Downarrow}_{V_E},(\val{\Downarrow}_{V_C},u),s')$. 

    A computation of $M_{w_0}$ corresponds to a computation of $C'$ that is projected on the states where $eval$ is true. Furthermore, $eval$ holds infinitely often along this computation. It follows that the computation of $C'$ satisifes $t(\phi)$ and we can conclude that the computation of $M_{w_0}$ satisfies $\phi$. 
\end{itemize}
\end{proof}

\subsubsection{Reactive Program Games}
We use the same partition of the set $V$ to environment and controller variables $V_E$ and $V_C$.
As before, we identify a finite set of predicates $P\subseteq \preds(V)$.
While in TSL we used individual updates for each $v\in V$, here, a global update is a tuple 
$\vec{u}=\langle u_v\rangle_{v\in V_C}$ including one update per output variable such that $u_v\in \mathcal{T}(V)$. Let $\vec{U}$ denote a finite set of global updates.

\begin{definition}[Reactive Program Game Structure]
    A reactive program game structure over $P$ and $\vec{U}$ is $\mathcal{G}=\langle V_E,V_C,L,l_0,Inv,\delta\rangle$, where $L$ is a finite set of locations, $l_0\in L$ is an initial location, $Inv:L\rightarrow \preds(V_C)$ maps each location to a location invariant, and $\delta \subseteq L\times P \times \vec{U} \times L$ is a transition relation.
    For every $l\in L$ the set $\delta(l)=\{ (l,p,\vec{u},l') \in \delta\}$ satisfies the following:
    \begin{itemize}
    \item
        $\bigvee_{(l,p,\vec{u},l')\in \delta(l)}p = \true$,
    \item 
        For every $(l,p_1,\vec{u}_1,l_1)$ and $(l,p_2,\vec{u}_2,l_2)$ such that $p_1\neq p_2$ we have $p_1\wedge p_2=\false$,
    \item  
        For every $(l,p,\vec{u},l_1)$ and $(l,p,\vec{u},l_2)$ we have $l_1=l_2$,
    \item 
        For every $x\in \Val(V_C)$ such that $x\models Inv(l)$ and for every $i\in\Val(V_E)$ there is some transition $(l,g,\vec{u},l')\in \delta(l)$ such that $(x\cup i) \models g$ and $\vec{u}(x\cup i)\models Inv(l')$.
    \end{itemize}
\end{definition}

Given an initial value $o_0\in\Val(V_c)$ such that $o_0\models Inv(l_0)$, a \emph{play} starting in $o_0$ of $\mathcal{G}$ is $(i_0,\vec{u}_i),(i_1,\vec{u}_1),\ldots$ such that for every $j\geq 0$ there is  
$(l_j,g_j,u_j,l_{j+1})\in \delta$ such that $(i_{j+1}\cup o_j)\models g_j$, $o_{j+1}=\vec{u}_j(i_{j+1}\cup o_j)$, and $o_{j+1}\models Inv(l_{j+1})$. 
We call the sequence $(o_0,l_0),\ldots$ the induced sequence of outputs and locations. 
We may refer to it implicitly when given a play and an initial value $o_0$. 

Given a program structure, we consider reachability, safety, and B\"uchi goals denoted by $\Omega\subseteq L$. 
A play satisfies the goal $\Omega$ if $\Omega$ is a reachability goal and for some $j$ we have $l_j\in \Omega$, if $\Omega$ is a safety goal and for all $j$ we have $l_j\in \Omega$, or if $\Omega$ is a B\"uchi goal and for infinitely many $j$ we have $l_j \in \Omega$. 

\begin{definition}[RPG Synthesis]
    A structure $\mathcal{G}$ and a goal $\Omega$ is realisable if and only if for every $o\in \Val(V_c)$ there is a Mealy machine $C_o$ with input $\Sigma_{in}=\Val(V_E)$ and output $\Sigma_{out}=\vec{U}$ such that every trace of $C_o$ is a play for the initial value $o$ that satisfies $\Omega$. The problem of synthesis is to construct the machines $\{C_o\}_{o\in \Val(V_C}\}$.
\end{definition}

Consider an RPG structure $\mathcal{G}$ over $P$ and $\vec{U}$, where all variables range over integers.
We define an arena that allows the environment to choose the initial valuation $o_0$ and in every step choose the valuation of the variables in $V_E$. This is done in a similar way to the way we handled TSL and explained above.
Let $A_{\mathcal G}$ be $\langle V_G,\val_0,\delta_G\rangle$, where $V_G=V\cup \{s,\ell\}\cup \mathbb{E}\cup \mathbb{C}$, where $\mathbb{E}=\{inc,dec,var_e,var_c\}$ as before, $\mathbb{C}$ is chosen such that $2^{\mathbb{C}}=\vec{U}$, $s$ ranges over $\{e,c\}$ as before, $\ell$ ranges over $L\cup\{err\}$, and $\val_0$ is an arbitrary valuation setting $s=c$ and $\ell=l_0$.
$$
\delta_G = \left \{ \left .
\begin{array}{r@{\quad \mapsto\quad }l}
(inc \wedge \neg dec) & v_{var_s}\inc\\
(\neg inc \wedge dec) & v_{var_s}\dec\\
(s=c \wedge (inc \iff dec) \wedge Inv(l_0) & s=e \\
(\ell = l \wedge \neg Inv(l) \wedge s=e \wedge (inc \iff dec) & \ell = err \\
(\ell = l \wedge Inv(l) \wedge s=e \wedge (inc \iff dec) \wedge g \wedge \vec{u} & \ell = l';\vec{u}
\end{array}
\right | (\ell,g,\vec{u},l')\in \delta  \right \} 
$$
Notice how we enforce that the environment sets the initial valuation of $V_C$ to a valuation that satisfies $Inv(l_0)$ and the new location $err$ is used to enforce that the controller chooses updates that satisfy the invariants of future locations. 

Given a goal, $\Omega$ we define $\phi_\Omega$ as follows.
By abuse of notation we relate to $\Omega$ as $\bigvee_{l\in \Omega} (\ell = l)$.
If $\Omega$ is a safety goal, we define $\phi_\Omega$ as $G \Omega$.
If $\Omega$ is a reachability goal, we define $\phi_\Omega$ as $(GF eval) \implies (F \Omega)$.
Finally, if $\Omega$ is a B\"uchi goal, we define $\phi_\Omega$ as $(GF eval) \implies (GF \Omega)$.

\begin{lemma}
    For every RPG $\mathcal{G}$ over $P$ and $\vec{U}$ and goal $\Omega$, $\mathcal{G}$ and $\Omega$ is realisable iff $\phi_\Omega$ is realisable modulo $A_{\mathcal G}$. 
\end{lemma}

The proof is very similar to the proof of Lemma~\ref{lem:tsl equirealisable} and is omitted.
\np{The translation from RPG to TSL is not going to be completed}
\end{toappendix}

\smallskip
\noindent
\textit{Infinite-state Arenas.}
Due to space restrictions,
we refer to other work~\cite{10.1145/3158149,10.1145/3632899} for a general overview of existing symbolic synthesis methods, and leave out infinite-state methods restricted to  decidable settings, such as pushdown games~\cite{DBLP:journals/iandc/Walukiewicz01}, Petri-net games~\cite{DBLP:conf/birthday/FinkbeinerO25}, or restrictions of FO-LTL such as those mentioned in the introduction~\cite{DBLP:conf/cav/RodriguezS23,DBLP:journals/jlap/RodriguezS24,DBLP:conf/aaai/Rodriguez024}.
Such approaches tend to apply very different techniques. 
We instead discuss methods that take on the undecidable setting, and how they acquire/encode liveness information. 
We find three classes of such approaches:

\smallskip
\noindent
\textit{Fixpoint solving.} These extend standard fixpoint approaches to symbolic game solving. 
\textsc{GenSys-LTL}~\cite{DBLP:journals/corr/abs-2306-02427} uses quantifier elimination to compute the controllable predecessor of a given set, terminating only if a finite number of steps is sufficient. A similar approach limits itself to the GR(1) setting~\cite{DBLP:conf/isola/MaderbacherWB24}, showing its efficiency also in the infinite setting. \texttt{rpgsolve}~\cite{10.1145/3632899} takes this further by finding so-called \emph{acceleration lemmas}. It attempts to find linear ranking functions with invariants to prove that loops in the game terminate, and thus it may find fixpoints that \textsc{GenSys-LTL} cannot. This information is however only used in a particular game region. In problems such as \texttt{robot-tasks}, this requires an infinite number of accelerations, leading to divergence.
The reliance on identifying one location in a game where a ranking function decreases
is also problematic when the choice of where to exit a region is part of the game-playing, or when the ranking needs to decrease differently based on the play's history.
{%
The latter would be required in order to scale their approach to objectives beyond B\"uchi and co-B\"uchi.}
The realisability solver \texttt{rpg-STeLA} tries to bypass the locality limitation by using game templates to identify lemmas that can be used in multiple regions. It does well on benchmarks that were designed for it in a compositional way, but in many other cases, the extra work required to identify templates adds significant overhead. For example, it causes divergence in \texttt{robot-tasks}. 
{%
As a bridge between program specifications in TSL and the \texttt{rpg} tools,} \texttt{tslmt2rpg}~\cite{DBLP:journals/pacmpl/HeimD25} translates TSL specifications to RPG while adding semantic information about infinite-range variables that allows it to simplify regions in games.
As for \texttt{rpg-STeLA} the analysis of the semantic information often causes a time overhead. Crucial here is the underlying solver, which often times out on quantifier elimination.
{%

\smallskip
\noindent
\textit{Abstraction.} Other methods, including ours, attempt synthesis on an explicit abstraction of the problem. 
A failure witness may be used to refine the abstraction and make another attempt. Some of these methods target games directly~\cite{DBLP:conf/icalp/HenzingerJM03,10.1007/978-3-540-74407-8_6,6987617}; others work at the level of the specification~\cite{10.1007/978-3-030-25540-4_35,MaderbacherBloem22,10.1145/3519939.3523429}. 
Many of these focus on refining states in the abstraction, a kind of safety refinement, 
as in the case of the tool \texttt{raboniel}~\cite{MaderbacherBloem22}. %
As far as we know, only \texttt{temos}~\cite{10.1145/3519939.3523429} adds some form of liveness information of the underlying infinite domain.
{%
It attempts to construct an abstraction of an \ltl (over theories) specification by adding consistency invariants, and transitions. It also uses syntax-guided synthesis to generate sequences of updates that force a certain state change}. Interestingly, it can also identify liveness constraints that abstract the effects in the limit of repeating an update $u$, adding constraints of the form $G(\textit{pre} \wedge (u W \textit{post}) \mathord{\implies} F \textit{post})$. However, it can only deal with one update of one variable at a time, and fails when the environment can delay $u$. Moreover, it does not engage in a CEGAR-loop, giving up if the first such abstraction is not realisable.

\smallskip
\noindent
\textit{Constraint Solving.} One may encode the synthesis problem into constrained Horn clauses (CHC), and synthesise ranking functions to prove termination of parts of a program. \texttt{Consynth}~\cite{DBLP:conf/popl/BeyeneCPR14} solves general \ltl and $\omega$-regular infinite-state games with constraint solving. However, it needs a controller template: essentially a partial solution to the problem. This may require synthesising ranking functions, and (unlike our approach) makes unrealisability verdicts limited to the given template and thus not generalisable.
MuVal~\cite{DBLP:journals/corr/abs-2007-03656} can encode realisability checking of \ltl games as validity checking in a fixpoint logic that extends CHC. It also requires encoding the automaton corresponding to the \ltl formula directly in the input formula, and discovers ranking functions based on templates to enforce bounded unfolding of recursive calls. Contrastingly, we do not rely on templates but can handle any argument for termination.

\section{Conclusions}\label{sec:conc}

We have presented a specialised CEGAR approach for \ltl synthesis beyond the Boolean domain. In our evaluation our implementation significantly outperforms other available synthesis tools, often synthesising a (counter-)strategy before other tools finish checking for realisability. Key to this approach are liveness refinements, which forgo the need for a large or infinite number of safety refinements. We carefully designed our framework so it can encode spuriousness checking of abstract \counterstrategies as simple invariant checking, using loops in counterexamples to find liveness refinements. Another main contribution is the reduction of the complexity of predicate abstraction and synthesis by an exponential, through a binary encoding of related predicates. This also allows to identify well-foundedness constraints of the arena, which we encode in the abstraction through LTL fairness requirements.

\smallskip\noindent
\emph{Future work.} 
We believe that symbolic approaches for LTL synthesis and synthesis for LTL over structured arenas~\cite{DBLP:conf/tacas/EhlersK24,DBLP:conf/fossacs/HausmannLP24}, could significantly benefit our technique. In these, determinisation for LTL properties would have to be applied only to the objective, and not to the arena abstraction. Tool support for these is not yet mature or available. For one such tool~\cite{DBLP:conf/tacas/EhlersK24}, we sometimes observed considerable speedup for realisability; however, it does not supply strategies.

Other directions include dealing with identified limitations (see Section~\ref{sec:eval}), extending the tool beyond LIA, dealing with infinite inputs automatedly, and applying other methods to manage the size of predicate abstractions, e.g.,~\cite{DBLP:conf/popl/HenzingerJMS02}, data-flow analysis, and implicit abstraction, and to make it more informative.

\clearpage
\bibliographystyle{splncs04}
\bibliography{references}

\renewcommand{\appendixsectionformat}[2]{Supplementary Material for Section~#1}

\end{document}